\documentclass[letterpaper,twocolumn,aps,prd,notitlepage,showkeys,showpacs,secnumarabic,citeautoscript,nobibnotes,floatfix,footinbib]{revtex4-1}
\usepackage[latin9]{inputenc}
\usepackage{array}
\usepackage{amsmath}
\usepackage{amssymb}
\usepackage{graphicx}
\usepackage{esint}

\makeatletter

\pdfpageheight\paperheight
\pdfpagewidth\paperwidth

\providecommand{\tabularnewline}{\\}

\usepackage{hyperref}

\AtBeginDocument{
  
}

\makeatother

\begin{document}

\title{Interplay between pair density waves and random field disorders in
the pseudogap regime of cuprate superconductors}
\begin{abstract}
To capture various experimental results in the pseudogap regime of
the underdoped cuprate superconductors for temperature $T<T^{*}$,
we propose a four-component pair density wave (PDW) state, in which
all components compete with each other. Without random field disorders
(RFD), only one of the PDW components survives. If the RFD is included,
this state could become phase separated and consist of short range
PDW stripes, in which two PDW components coexist but differ in magnitudes,
resulting in charge density waves (CDW) and a time-reversal symmetry
breaking order, in the form of loop current, as secondary composite
orders. We call this phase-separated pair nematic (PSPN) state, which
could be responsible for the pseudogap. Using a phenomenological Ginzburg-Landau
approach and Monte Carlo simulations, we found that in this state,
RFD induces short range static CDW with phase-separated patterns in
the directional components and the static CDW is destroyed by thermal
phase fluctuations at a crossover temperature $T_{CO}<T^{*}$, above
which the CDW becomes dynamically fluctuating. The experimentally
found CDW with predominantly $d$-wave form factor constrains the
PDW components to have $s^{\prime}\pm id$ pairing symmetries. We
also construct a lattice model and compute the spectral functions
for the PSPN state and find good agreement with ARPES results. 
\end{abstract}

\author{Cheung Chan}

\affiliation{Institute for Advanced Study, Tsinghua University, Beijing, 100084,
China}


\maketitle

\section{Introduction}

Since the discovery of the cuprate superconductors \cite{Bednorz1986}
(SC), the pseudogap regime has attracted a lot of attentions for its
exotic properties \cite{Orenstein21042000} and the possible relation
to the high temperature $d$-wave superconductivity (dSC). Two main
types of theories are proposed for the pseudogap. One type \cite{Emery1995,Franz1998,Lee2006,Perali2002}
suggests that pseudogap is a precursor to the dSC phase with pre-formed
Cooper pairs without global phase coherence. However, mounting experimental
evidences are pointing otherwise that pseudogap is indeed another
broken symmetry state. The recent inputs mainly come from the X-ray
scattering experiments \cite{Ghiringhelli17082012,Chang2012,PhysRevLett.109.167001,PhysRevLett.110.137004,PhysRevLett.110.187001,Comin24012014,Comin20032015,DaSilvaNeto2014,2014arXiv1402.5415C},
which suggest the state is characterized by the onset of incommensurate
charge density wave (CDW) with wave vectors at $(\pm2Q,0)$ and $(0,\pm2Q)$,
where $Q$ decreases with increasing hole doping. This finding is
in accord with earlier scanning tunneling microscope \cite{Wise2008,Parker2010,Lawler2010,Mesaros22072011,Fujita2014}
(STM) and nuclear magnetic resonance \cite{Wu2011,Wu2015} (NMR) experiments.
There are some early theoretical studies to explain the pseudogap
in terms of CDW \cite{Kivelson1998,Bianconi1998}. The phase diagram
is however much richer. Other orders associated with different broken
symmetries are also detected in the pseudogap, e.g., the time reversal
symmetry breaking (TRSB) order detected in polar Kerr rotation \cite{Xia2008,Karapetyan2014}
(PKR) and polarized neutron diffraction \cite{Fauque2006,Li2008,Mangin-Thro2015,PhysRevLett.105.027004}
(PND) experiments and the nematic order \cite{Hinkov2007,Daou2010,Lubashevsky2014,Comin20032015,Cyr-Choiniere2015}
that breaks the $C_{4}$ lattice rotational symmetry. This complicated
zoo of orders poses a natural question that whether we can unify these
into a common origin and understand the pseudogap in a coherent manner.
Among the various theories proposed, pair density wave (PDW) order,
which is a spatially modulating SC state similar to Fulde-Ferrell-Larkin-Ovchinnikov
(FFLO) states \cite{Fulde1964,Larkin1965}, previously studied in
different contexts for the cuprates \cite{Himeda2002,Kivelson2003,Chen2004,Berg2007,Agterberg2008,Seo2008,Berg2009,Berg2009a,Berg2009b,RevModPhys.87.457},
is suggested to be responsible for the pseudogap. Recent works \cite{PhysRevX.4.031017,Agterberg2015}
by Lee and Agterberg \textit{et\,al}.\,have successfully explained
many features of the pseudogap. In the PDW theory, the CDW is induced
by the PDW as a secondary composite order, accounting for the STM
and X-ray scattering results. Moreover, the PDW can also induce a
loop current (LC) order \cite{Simon2002,Varma2014,Agterberg2015}
to account for the TRSB order observed in PKR \cite{Xia2008,Karapetyan2014}
and PND \cite{Fauque2006,Li2008,PhysRevLett.105.027004,Mangin-Thro2015}
experiments. Nonetheless, the PDW order also explain several ARPES
features \cite{Norman1998,Comin24012014,Lee2007a,Kondo2009,Hashimoto2010,He25032011},
namely the $k_{F}$-$k_{G}$ misalignment, antinodal gap closing from
below, and the Fermi arcs.

Hinted by these successes, the PDW order might hold the key to understand
the pseudogap of hole doped cuprates. However, several issues have
surfaced. Most noticeably, the charge density waves observed in STM
\cite{Mesaros22072011,Fujita2014,Hamidian2015} and REXS \cite{Comin20032015}
are short ranged. STM further show that these density waves form a
domain structure and are directional within each domain. The same
conclusion is also drawn from REXS. While these results are obtained
in no magnetic field, the high field experiments, on the other hand,
reveal that the CDW with the same in-plane wave vector is long ranged
and unidirectional \cite{Gerber2015,2015arXiv150805486G,2015arXiv151106092C,Wu2011}.
 As a first step to understand this onset of long range CDW state
at high field \cite{Wu2015,2015arXiv150805486G,2015arXiv151106092C},
it is essential to explain why the CDW observed in weak field is short
ranged.

Experiments however impose stronger constraints, besides the possible
order parameters, on theories. Each set of experiments reveals a specific
doping dependent temperature scale but not all these temperature scales
can fall into a single simple curve in the phase diagram. This challenges
the belief that the pseudogap can be ascribed to a single origin.
We shall mention those that are relevant in the current paper. The
electronic transport giving rise to linear resistivity \cite{PhysRevLett.93.267001,Daou2009}
defines $T_{\rho}$ that is believed to capture a quantum critical
point associated with corresponding order parameter and it is commonly
taken as the definition of the pseudogap temperature scale $T^{*}\equiv T_{\rho}$.
ARPES measurements define $T_{\mathrm{ARPES}}\sim T^{*}$, below which
unusual quasiparticle spectrum is revealed. Moreover, REXS \cite{Ghiringhelli17082012,PhysRevLett.109.167001,PhysRevLett.110.187001,Comin24012014,Comin20032015,DaSilvaNeto2014,2014arXiv1402.5415C}
also found the fluctuating CDW correlation below $T^{*}$, but RIXS
\cite{Chang2012,PhysRevLett.110.137004} and STM \cite{Parker2010}
reveal a lower temperature scale $T_{CO}<T^{*}$ for static charge
order. It is found that $T_{CO}$ has a maximum at some doping but
$T^{*}$ decreases as doping increases. A valid theory would be able
to explain the emergence of these temperature scales within the pseudogap.
Similar to $T_{CO}$, we note that for TRSB order, there are also
two distinct temperature scales found in PND \cite{Fauque2006,Li2008,Mangin-Thro2015}
(see also Ref.\,\cite{PhysRevLett.105.027004}) and PKR \cite{Xia2008,He25032011}
(see also Ref.\,\cite{Karapetyan2014}) experiments. The PND experiments
found that below $T_{M}\sim T^{*}$ an intra-unit cell (IUC) TRSB
magnetic order exists, while the PKR experiments, which also detect
TRSB, reveal a strictly lower temperature scale $T_{K}<T_{M}$ but
doping dependence of $T_{K}$ shows a similar trend as $T_{M}$. However,
we shall address the issue of $T_{K}$ elsewhere.

\begin{figure}
\begin{centering}
\includegraphics[width=0.65\columnwidth]{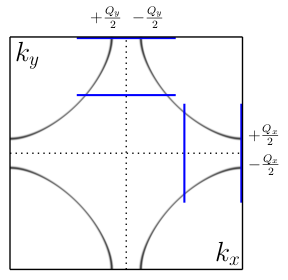}
\par\end{centering}

\caption{\label{fig:BZ}(Color online) Bare Fermi surface. The PDW pairing
centers $\pm Q_{x,y}/2$ are at the intersections with the Brillouin
zone boundary. The blue lines are for the quasiparticle spectral function
scans in Fig.\,\ref{fig:PDW2-ARPES} and \ref{fig:PDW4-ARPES}.}
\end{figure}

\begin{figure}
\includegraphics[width=0.7\columnwidth]{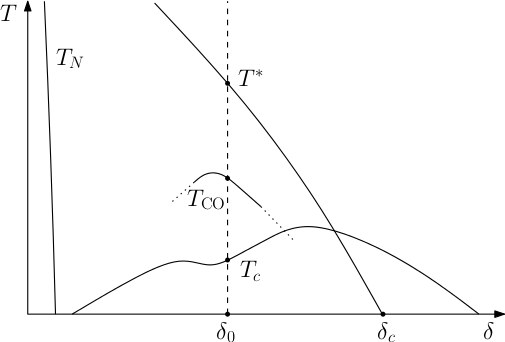}

\caption{\label{fig:phaseDiag}Schematic phase diagram for hole doped cuprates,
where $\delta$ is the doping and $T$ is temperature. $T^{*}$ is
the characteristic temperature for the pseudogap, in which we propose
the phase-separated pair nematic state is responsible for. $T_{N}$
is the critical temperature for the antiferromagnetic order near zero
doping. Focusing at a doping $\delta_{0}$ at underdoped regime, $T_{c}$
is the critical temperature for the dSC, $T_{CO}$ is for the presence
of short range static CDW.}
\end{figure}

Motivated by these issues, we study the effect of random field disorders
\cite{Larkin1970,Imry1975} (RFD) in a four-component PDW (PDW4) model
\cite{Berg2009a,PhysRevX.4.031017,RevModPhys.87.457} (see also Ref.\,\cite{Nie2014}
for the study of a CDW model). Intuitively, RFD does nothing more
than inducing the PDW short range \cite{PhysRevX.5.031008}. But we
will show below that in order to explain the experimental results
coherently, inclusion of RFD would lead us to consider a PDW state
of different nature unexplored before, thus it is not straightforward
to generalize existing PDW results to our case. We consider four PDW
components $\Delta_{\pm Q_{x,y}}$ at wave vectors $\pm Q_{x}$ and
$\pm Q_{y}$ and all the components compete with each other. The pairing
centers $\pm Q_{x}/2$ and $\pm Q_{y}/2$ are located at the intersections
of the Fermi surface and the Brillouin zone boundary as shown in Fig.\,\ref{fig:BZ}.
We imagine that the PDW orders present below a characteristic temperature
$T^{*}$ that decreases as doping increases, as shown in Fig.\,\ref{fig:phaseDiag}.
$T^{*}$ terminates \textit{inside} the dSC phase at a quantum critical
point $\delta_{c}$, which controls a quantum critical region \cite{Sachdev2000,Valla1999,Marel2003}
with special properties like linear electronic resistivity \cite{PhysRevLett.93.267001}
(see also Ref.\,\cite{Daou2009}). We first use Ginzburg-Landau (GL)
formulation and Monte Carlo simulations to study the interplay of
PDW, CDW and RFD in real space. A PDW with $\pm Q$ components would
induce a secondary composite CDW order with wave vector $2Q$ \cite{Agterberg2008,Berg2009a,Berg2009b,RevModPhys.87.457},
which is then coupled to the RFD \cite{PhysRevX.5.031008}. With the
$\pm Q$ PDW components compete with each other and the presence of
RFD, it is possible to induce a TRSB loop current order compatible
with the CDW. We also show that there exists a crossover temperature
scale $T_{CO}$ (see Fig.\,\ref{fig:phaseDiag}) for the short range
static CDW order and argue that the PDW superconductivity is absent
due to SC phase fluctuations. Moreover, RFD induces a domain pattern,
or phase separation, on the directional PDW components at $\pm Q_{x}$
and $\pm Q_{y}$. We shall dub this state of short range PDW with
four competing components the phase-separated pair nematic (PSPN)
state. Next, we argue that the predominantly $d$-wave form factor
CDW observed in STM constrains the PDW to be a bond order with $s'\pm id$-wave
pairing symmetry. Using this input together with a lattice model and
the GL functional, we compute the quasiparticle spectral functions
under the influence of thermal fluctuations and find good agreement
with ARPES.

In this study, we make several assumptions. We consider only the cuprate
families that show CDW of decreasing wave vectors with increasing
doping, as seen in recent X-ray scattering experiments \cite{PhysRevLett.110.137004,DaSilvaNeto2014,Comin24012014}.
In this study, we have neglected the cuprate families like LSCO and
LBCO that show CDW wave vectors scaling with doping \cite{Kivelson2003}
and the CDW form factor appears to be $s'$-wave \cite{2014arXiv1409.6787A},
though PDW has also been suggested to play a role \cite{Berg2007,Berg2009,Berg2009a,PhysRevX.5.031008}.
We shall focus on the experimental results in weak magnetic fields,
because some recent experiments \cite{Gerber2015,2015arXiv150805486G,2015arXiv151106092C,Wu2015,Wu2011}
reveal that the CDW order observed in high field and low temperature
could be of different nature than the one obtained in weak field.
In the following, we neglect the dSC for simplicity and only consider
the PDW order and its possible induced composite orders at a specific
doping $\delta_{0}$ in the phase diagram (Fig.\,\ref{fig:phaseDiag}).
It is suggested that the $d$-wave superconductivity competes with
the CDW \cite{Efetov2013,Hayward21032014,Achkar2014,Campi2015}. Lastly,
the origin of the PDW in cuprate is not the focus of the present paper,
but some previous studies show that PDW could be originated from strong
correlations \cite{Corboz2014,PhysRevX.4.031017}.

There is another line of theory that consists PDW as an important
player. Its starting point is a spin-fluctuation scenario, captured
within the spin-fermion model \cite{Metlitski2010a}. PDW and CDW
with $d$-wave form factors can emerge naturally from this semi-microscopic
model \cite{Pepin2014,Wang2015,Wang2015a,Freire2015} (see also Ref.\,\cite{Perali1996})
and it can also capture a number of important properties and features
observed in the experiments. In particular, an unidirectional CDW/PDW
mixed state \cite{Wang2015,Wang2015a} obtained from the model can
account for the ARPES spectrum, TRSB order, CDW order, and breaking
of $C_{4}$ symmetry. Although these results are similar to those
of the PSPN state we discuss here, we note that in the spin-fermion
model, PDW and CDW orders form a ``supervector'' stemming from an
approximate SU(2) symmetry near the intersection points of the Fermi
surface and magnetic Brillouin zone (the ``hot-spots''), while here
the PSPN state has the PDW as the main character and the CDW plays
only a parasitic role \cite{PhysRevX.4.031017,Agterberg2015}. As
we will show below, this parasitic CDW plays a crucial part in explaining
the distinct temperature scales observed in the pseudogap, namely
$T^{*}$ and $T_{CO}$. Here we wish to consider the parasitic CDW
induced by the PDW only and see how far this perspective can proceed.

The paper is outlined as follow: In section \ref{sec:PDW2}, we will
first briefly review the two-component PDW state, and followed by
the discussion of the interplay between the four-component PDW state
and RFD in section \ref{sec:PDW4}. In section \ref{sec:Constraint},
we constrain the PDW pairing symmetries from the CDW experiments.
In section \ref{sec:ARPES}, we present the results for the quasiparticle
spectral functions and compare with the ARPES experiments. Due to
the number of acronyms, we provide an acronym list for the phrases
commonly used at the end of the paper to facilitate the reading.

\section{Brief review on Two-Component PDW (PDW2)\label{sec:PDW2}}

In this section, we briefly review and discuss the model of two-component
PDW (PDW2) and the effect of RFD. The main focus of the paper is to
study the PDW with four components and relate to the experimental
results in the pseudogap regime of the cuprates. In order to do that,
it is beneficial to first discuss the PDW2 system \cite{Kivelson2003,Berg2007,Berg2009,Berg2009a,Berg2009b,RevModPhys.87.457,PhysRevX.5.031008},
from which we can easily generalize the results to the four component
case. Firstly, we will introduce the PDW order parameters consisting
two components and the corresponding GL functional. Also it is possible
to induce a CDW from the PDW2, provided that both of the PDW components
are non-vanishing. Next, we imagine that the CDW are coupled to the
charged impurities, which are modeled by the coupling between CDW
and RFD in the GL functional. For a moment, we shall consider only
the subproblem of the CDW coupled to RFD, which can be cast into the
random field XY model \cite{Gingras1996,PhysRevLett.82.1935}. (Actually
the random field XY model and variants have been taken as minimal
models of a broad class of condensed matter systems with quenched
disorders, such as flux-line arrays in dirty type-II superconductors
\cite{Blatter1994}, charge density waves \cite{Gruner1988}, and
smectic liquid crystals in random environments \cite{Radzihovsky1997,Radzihovsky1997a,Radzihovsky1999})
In the model, the XY phases (CDW smectic phase) are coupled to the
RFD. It is found \cite{Gingras1996,PhysRevLett.82.1935} that in 2D
the RFD, no matter how weak, always induces topological defects at
long scales, leading to short range (exponential) correlations in
the CDW. Without disorder, the model reduces to a pure XY model and
it allows a Berezinsky-Kosterlitz-Thouless (BKT) phase transition
\cite{Berezinskii1971,KosterlitzJ.M.;Thouless1973} from a low temperature
phase with power law correlation to a high temperature phase with
exponential correlation. With finite RFD, the low temperature ``phase''
readily has exponential correlation, and the residue of the phase
transition is a crossover \cite{Gingras1996}, below which only short
range \textit{static} CDW order exists. Now we return to the full
model containing PDW2, CDW and RFD, which has been studied before
in Ref.\,\cite{PhysRevX.5.031008}. From the results of random field
XY model, we readily conclude that the CDW has exponential correlation.
This also implies the absence of long range PDW order, since long
range PDW would have induced CDW with long range correlation, contradicting
the fact that the system with RFD only has short range CDW. Besides,
we also expect a crossover temperature for the short range static
CDW at low temperature. To this end, we will review the topological
defects \cite{Berg2009b,Radzihovsky2009,Radzihovsky2011,Barci2011}
arising in the PDW2 system (section \ref{sub:Topo-defects}) and the
RFD induced crossover (section \ref{sub:PDW2-with-RFD}). We also
present numerical evidence to show the presence of these topological
defects in the PDW2 system with finite RFD in section \ref{sub:Numerics-PDW2}
and lay the foundation for later numerical studies of the four-component
PDW systems.

We first consider PDW2 state and its order parameter, which generally
is 
\begin{equation}
\Delta(R)=\Delta_{+Q}(R)e^{+iQ\cdot R}+\Delta_{-Q}(R)e^{-iQ\cdot R}\thinspace,
\end{equation}
where $\Delta_{\pm Q}(R)\in\mathbb{C}$. In particular, if $\Delta_{\pm Q}(R)=\Delta_{0}/2$
are equal and homogeneous in space, then the PDW order $\Delta(R)=\Delta_{0}\cos Q\cdot R$
shows modulation of wave vector $Q$ in real space. This state induces
a CDW order at wave vector $2Q$, such that $\rho_{2Q}(R)\sim\Delta_{-Q}^{*}(R)\Delta_{+Q}(R)$.

In order to discuss the influence of random field disorders on the
PDW and CDW state, we write down the Ginzburg-Landau (GL) functional
(density) for PDW2 \cite{PhysRevX.5.031008,RevModPhys.87.457}:
\begin{eqnarray}
F_{2}^{\Delta} & = & +\alpha^{\Delta}\left|\Delta\right|^{2}+\gamma_{1}^{\Delta}\left|\Delta\right|^{4}+\gamma_{3}^{\Delta}\left|\Delta_{+Q}\right|^{2}\left|\Delta_{-Q}\right|^{2}\nonumber \\
 &  & +\kappa_{\perp}^{\Delta}\left|\partial_{x}\Delta\right|^{2}+\kappa_{\parallel}^{\Delta}\left|\partial_{y}\Delta\right|^{2}\thinspace,
\end{eqnarray}
where $\Delta=\left(\Delta_{+Q},\Delta_{-Q}\right)^{T}$. In general,
the system could admit nematicity and $\kappa_{\parallel,\perp}^{\Delta}$
would differ. The other part of the GL functional involves the CDW
and is given by
\begin{equation}
F_{2}^{\rho}=-H^{*}\rho_{2Q}+\mathrm{c.c.}+\kappa_{\perp}^{\rho}\left|\partial_{x}\rho_{2Q}\right|^{2}+\kappa_{\parallel}^{\rho}\left|\partial_{y}\rho_{2Q}\right|^{2}\thinspace,\label{eq:F2b}
\end{equation}
where $H(R)=h_{\mathrm{rms}}h(R)e^{i\eta(R)}$ is the local complex
random field, $h(R)$ are Gaussian-distributed random variables with
mean $0$ and standard deviation $1$, $\eta(R)\in[0,2\pi)$ are uniformly
distributed random phases, and $h_{\mathrm{rms}}\geqslant0$ controls
the (root mean square) random field disorder strength. The random
field term models randomly distributed non-magnetic charged impurities
that pin down the CDW modulations. Notice that the RFD couples to
the CDW, which reflects the charge modulations of the PDW, but not
directly to the PDW. In the following, we assume that the PDW always
induces a CDW $\rho(R)\sim\rho_{2Q}\cos2Q\cdot R$ such that 
\begin{equation}
\rho_{2Q}=\Delta_{-Q}^{*}\Delta_{+Q}\thinspace.
\end{equation}
In this case, the total GL functional can be expressed in terms of
$\Delta_{\pm Q}$ only as
\begin{eqnarray}
F_{2} & = & F_{2}^{\Delta}+F_{2}^{\rho}\\
F_{2}^{\rho} & = & -H^{*}\Delta_{-Q}^{*}\Delta_{+Q}+\mathrm{c.c.}\nonumber \\
 &  & +\kappa_{\perp}^{\rho}\left|\partial_{x}\left(\Delta_{-Q}^{*}\Delta_{+Q}\right)\right|^{2}+\kappa_{\parallel}^{\rho}\left|\partial_{y}\left(\Delta_{-Q}^{*}\Delta_{+Q}\right)\right|^{2}\thinspace.\nonumber 
\end{eqnarray}
We note that $\gamma_{3}^{\Delta}$ controls whether $\Delta_{\pm Q}$
components compete with each other. In this section, we shall consider
$\gamma_{3}^{\Delta}<0$ such that two $\Delta_{\pm Q}$ components
coexist and induce a $2Q$-CDW, even in the absence of random field
disorder $H(r)=0$. The case $\gamma_{3}^{\Delta}>0$ will be discussed
in section \ref{sec:PDW4}.

\subsection{Topological defects\label{sub:Topo-defects}}

Here we briefly discuss the possible topological defects \cite{Agterberg2008,Berg2009b,Radzihovsky2009,Radzihovsky2011}
in the PDW2 system (in the absence of RFD). Assuming $\gamma_{3}^{\Delta}<0$,
then we have $|\Delta_{+Q}|\approx|\Delta_{-Q}|$. We thus write $\Delta_{\pm Q}(R)=\frac{1}{2}\Delta_{0}e^{i\theta_{\pm Q}(R)}$,
where we assume the amplitude fluctuations are small compared to those
of the phases $\theta_{\pm Q}$. The PDW order is now
\begin{equation}
\Delta(R)=\Delta_{0}e^{i\vartheta}\cos\left(Q\cdot R+\varphi\right)\thinspace,
\end{equation}
where the superconducting phase is $\vartheta=\frac{1}{2}\left(\theta_{+Q}+\theta_{-Q}\right)$
and the PDW smectic phase is $\varphi=\frac{1}{2}\left(\theta_{+Q}-\theta_{-Q}\right)$.
Deep in the PDW2 phase in 2D, the amplitude fluctuations are negligible,
the system can be effectively described by an anisotropic XY model
in terms of the $\vartheta$ and $\varphi$ \cite{Berg2009b,Radzihovsky2009,Radzihovsky2011,Barci2011}.
The PDW2 system possesses three types of topological defects labeled
by $(n_{v},n_{d})$: (1) pure SC $2\pi$-vortices with $(\pm1,0)$,
(2) $2\pi$-dislocations with $(0,\pm1)$, and $\pi$-vortex-$\pi$-dislocation
defect, or simply half-vortex, with $(\pm\frac{1}{2},\pm\frac{1}{2})$.
The labels $(n_{v},n_{d})$ are related to the topological singularities
(topological ``charges'') in $\vartheta(R)$ and $\varphi(R)$,
defined by
\begin{eqnarray}
\oint d\vec{\ell}\cdot\nabla\vartheta & = & 2\pi n_{v},\nonumber \\
\oint d\vec{\ell}\cdot\nabla\varphi & = & 2\pi n_{d}\thinspace.
\end{eqnarray}
Due to the single-value conditions in the original $\theta_{\pm Q}(r)$
fields, $n_{v,d}$ can take half-integer or integer values and it
leads to three types of defects in the system. Moreover, starting
with a PDW2 at low temperature, owing to (thermal) proliferation of
different defects, the system restores broken symmetries and yields
various phases \cite{Berg2009b,Radzihovsky2011,Barci2011}. A pure
CDW phase is accessed when the SC $2\pi$-vortices are proliferated
and a charge-4e SC phase is obtained through the proliferation of
$2\pi$-dislocations. To attain a nematic phase one needs to proliferate
the half-vortices. We note that since the PDW order modulates half
as many as its induced CDW, the PDW half-vortex, manifested as a $2\pi$-dislocation
in the CDW, is accompanied by a half SC flux $\Phi_{0}/2$, which
can be experimentally verified \cite{PhysRevX.5.031008}.

\subsection{PDW2 with RFD\label{sub:PDW2-with-RFD}}

It is well known that the random field disorder destroys CDW long
range order for dimension $d\leq4$ \cite{Larkin1970,Imry1975}. Consider
the RFD term $-H^{*}\Delta_{-Q}^{*}\Delta_{+Q}+\mathrm{c.c.}\propto-\cos\left(\eta-2\varphi\right)$
in Eq.\,\eqref{eq:F2b}, the CDW smectic phase $2\varphi$ tends
to align with the RFD phase $\eta$. Treating the random field as
a perturbation, the energy gain due to disorder potential $\sim L^{d/2}$
dominates over the elastic energy cost of adjusting to disorder $\sim L^{d-2}$.
As a result, long range CDW is inhibited beyond a length scale $\xi_{L}$,
known as the Larkin length. This argument implies the absence of long
range PDW order if the CDW is induced by the PDW.

This picture only considers continuous elastic deformation of the
uniform state, but not topological defects, which are non-perturbative
in nature. Indeed, via numerical studies \cite{Gingras1996,PhysRevLett.82.1935},
it is shown that these topological defects always exist for any RFD
strength in 2D. In the absence of these defects, the CDW correlation
decays in power law, admitting a genuine phase transition to a high
temperature phase with exponentially decaying correlation. In 2D,
however, at a length scale $\xi_{V}>\xi_{L}$, the equilibrium system
is always unstable to a proliferation of the static topological defects
and thus the CDW correlation decays exponentially for any RFD strength.

In 2D, these CDW topological defects manifest as $2\pi$-dislocations.
Re-tracking back to the PDW, it means a PDW half-vortex exists exactly
at the CDW $2\pi$-dislocation \cite{PhysRevX.5.031008}. Proliferation
of these half-vortices destroys the long range PDW and lead to a nematic
phase. These topological defects are pinned and static (over thermal
average) at low temperature, while they can diffuse driven by thermal
fluctuations at higher temperature. The presence of these defects
results in short range correlations in PDW and CDW \cite{PhysRevX.5.031008}.
Instead of a genuine phase transition, we thus expect a crossover
temperature $T_{CO}$, below which short range static charge order
can be observed, and above which most of the charge order is destroyed
by the defect diffusion (except domains that are pinned strongly by
the RFD). Here only an argument is given, we shall provide numerical
evidence in section \ref{sec:PDW4}.

\subsection{Numerics: Monte Carlo study of PDW2 with RFD\label{sub:Numerics-PDW2}}

\begin{figure}
\begin{centering}
\includegraphics[width=0.75\columnwidth]{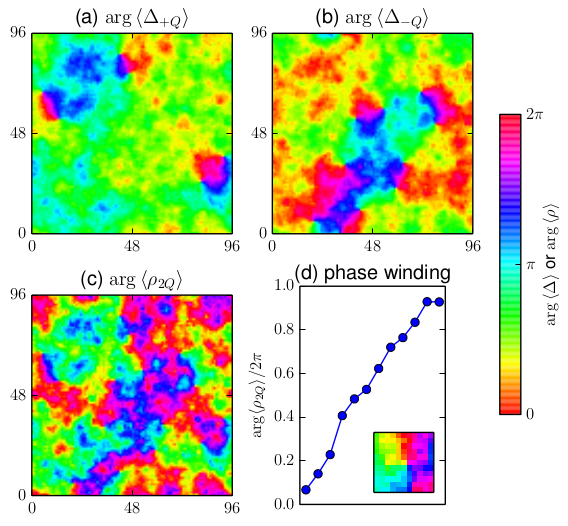}
\par\end{centering}

\caption{\label{fig:pdw2-cdw}(Color online) Monte Carlo results for PDW2 system
in $F_{2,\mathrm{lattice}}$ {[}Eq.\,\eqref{eq:F_2}{]}. The phases
of the average orders (a) $\left\langle \Delta_{+Q}\right\rangle $,
(b) $\left\langle \Delta_{-Q}\right\rangle $, and (c) $\left\langle \rho_{2Q}\right\rangle $
are shown. Here we choose $\gamma_{3}^{\Delta}<0$ such that we can
consider the phases and their fluctuations only and assume constant
order magnitudes. In (a) and (b), several defect pairs with opposite
windings are present in $\left\langle \Delta_{\pm Q}\right\rangle $.
Due to the fact that the CDW is induced by the PDW, these defect pairs
also manifest in the $\left\langle \rho_{2Q}\right\rangle $ {[}see
(c){]}. As an example, panel (d) shows a $2\pi$ CDW smectic phase
winding around a counterclockwise path of a CDW $2\pi$-dislocation
(see the insert) located near $\vec{R}=(53,60)$. Parameters (in meV):
$\kappa_{\parallel,\perp}^{\Delta}=5$, $\kappa_{\parallel,\perp}^{\rho}=1$,
$h_{\mathrm{rms}}=4$, $\alpha^{\Delta},\gamma_{3}^{\Delta}<0$ and
$\gamma_{1}^{\Delta}>0$. System size: $96\times96$ with periodic
boundary condition. Simulated annealing is performed from a high temperature
to the final temperature at $10$\,K.}
\end{figure}

In order to gain more intuition on the interplay of PDW, CDW and RFD
in a PDW2 system, we perform Monte Carlo study on the corresponding
lattice GL functional \cite{PhysRevX.5.031008}, 
\begin{equation}
F_{2,\mathrm{lattice}}=F_{2,\mathrm{lattice}}^{\Delta}+F_{2,\mathrm{lattice}}^{\rho}\label{eq:F_2}
\end{equation}
 with 
\begin{eqnarray*}
 &  & F_{2,\mathrm{lattice}}^{\Delta}=\\
 &  & +\tilde{\alpha}^{\Delta}\sum_{\mathbf{i}s}\left|\Delta_{sQ}(\mathbf{i})\right|^{2}+\gamma_{1}^{\Delta}\sum_{\mathbf{i}}\left(\sum_{s}\left|\Delta_{sQ}(\mathbf{i})\right|^{2}\right)^{2}\\
 &  & +\gamma_{3}^{\Delta}\sum_{\mathbf{i}}\left|\Delta_{+Q}(\mathbf{i})\right|^{2}\left|\Delta_{-Q}(\mathbf{i})\right|^{2}\\
 &  & -2\kappa_{\perp}^{\Delta}\sum_{\mathbf{i}s}\mathrm{Re}\left[\Delta_{sQ}(\mathbf{i}+\hat{x})\Delta_{sQ}^{*}(\mathbf{i})\right]\\
 &  & -2\kappa_{\parallel}^{\Delta}\sum_{\mathbf{i}s}\mathrm{Re}\left[\Delta_{sQ}(\mathbf{i}+\hat{y})\Delta_{sQ}^{*}(\mathbf{i})\right]
\end{eqnarray*}
\begin{eqnarray*}
 &  & F_{2,\mathrm{lattice}}^{\rho}=\\
 &  & -2h_{\mathrm{rms}}\sum_{\mathbf{i}}\mathrm{Re}\left[h(\mathbf{i})e^{-i\eta(\mathbf{i})}\Delta_{-Q}^{*}(\mathbf{i})\Delta_{+Q}(\mathbf{i})\right]\\
 &  & -2\kappa_{\perp}^{\rho}\sum_{\mathbf{i}}\text{Re}\left[\Delta_{-Q}(\mathbf{i}+\hat{x})\Delta_{+Q}^{*}(\mathbf{i}+\hat{x})\Delta_{-Q}^{*}(\mathbf{i})\Delta_{+Q}(\mathbf{i})\right]\\
 &  & -2\kappa_{\parallel}^{\rho}\sum_{\mathbf{i}}\text{Re}\left[\Delta_{-Q}(\mathbf{i}+\hat{y})\Delta_{+Q}^{*}(\mathbf{i}+\hat{y})\Delta_{-Q}^{*}(\mathbf{i})\Delta_{+Q}(\mathbf{i})\right]\\
 &  & +2\left(\kappa_{\parallel}^{\rho}+\kappa_{\perp}^{\rho}\right)\sum_{\mathbf{i}}\left|\Delta_{-Q}(\mathbf{i})\Delta_{+Q}(\mathbf{i})\right|^{2}
\end{eqnarray*}
on a square lattice. Here $\tilde{\alpha}^{\Delta}=\alpha^{\Delta}+2\kappa_{\parallel}^{\Delta}+2\kappa_{\perp}^{\Delta}$
and the summation is over site $\mathbf{i}$ and sign $s=\pm$. $\hat{x}$
and $\hat{y}$ are shifts by one lattice constant in the corresponding
directions in real space. The random fields $h(\mathbf{i})e^{i\eta(\mathbf{i})}$
are Gaussian-distributed at each site. If we choose $\gamma_{3}^{\Delta}<0$,
we can safely set the amplitudes $\left|\Delta_{\pm Q}\right|$ as
constants for the system at low enough temperature and only concern
about the phase fields $\theta_{\pm Q}(\mathbf{i})$. This essentially
reduces to a random field XY model \cite{PhysRevX.5.031008}, on which
we perform the Monte Carlo (MC) simulations with Metropolis algorithm
of single-site phase rotations. The initial PDW phases $\theta_{\pm Q}$
are initialized randomly at a high enough temperature, which is then
slowly lowered to the desired one. This simulated annealing process
lets the system to be better equilibrated at the final temperature.
At the final temperature, we measure the thermal averaged phases of
$\Delta_{\pm Q}$ and $\rho_{2Q}=\Delta_{-Q}^{*}\Delta_{+Q}$ at each
site. For simplicity, we choose $\kappa_{\parallel}=\kappa_{\perp}$.

In Fig.\,\ref{fig:pdw2-cdw}, the results are shown for a typical
RFD configuration. Due to their single-value conditions, we show the
$\theta_{\pm Q}$ fields instead of the $\vartheta$ and $\varphi$
fields. The choice of parameters deserves some comments. Restricting
to $\gamma_{3}^{\Delta}<0$ (i.e.\,coexisting $\Delta_{\pm Q}$),
the phase diagram for $F_{2,\mathrm{lattice}}$ turns out to be simple.
In Ref.\,\cite{PhysRevX.5.031008} it is found that there are only
two phases: the fully disordered phase with the presence of all types
of topological defects, and the PDW glass phase, in which no CDW dislocation
of $2\pi$-winding is allowed (only $4\pi$-defects are permitted).
The parameters in Fig.\,\ref{fig:pdw2-cdw} correspond to the fully
disordered phase due to the presence of $2\pi$-defects. The PDW glass
phase is interesting in its own right, but we shall only consider
the fully disordered phase here. Back to Fig.\,\ref{fig:pdw2-cdw},
several observations are in order. First, the CDW phase $2\varphi(\mathbf{i})$
forms ``domains.'' This is consistent with our previous discussion
that a few sites bare strong random fields and the CDW phases near
these sites would align along with these strong random fields under
the influence of the stiffness terms $\kappa^{\Delta,\rho}$. Next
is the existence of $2\pi$-dislocations ($2\pi$-windings) in the
CDW phase {[}Fig.\,\ref{fig:pdw2-cdw}(c) and (d){]}. We note that
these defects occur near the domain boundary since doing so the system
can lower its energy better than pure continuous elastic deformation
\cite{Gingras1996}. These dislocations come in pairs of opposite
windings, which is expected in the system with periodic boundary condition
and zero total winding.

We note that this simulated method can be utilized to find the ground
state of the system, but what we have found might not be the true
ground state and it is not our main object to do so. Indeed, the system
shows glassy behavior that it can be trapped in some local energy
minimum. This can be understood in terms of the phase relation $2\varphi=\theta_{+Q}-\theta_{-Q}$
that although the CDW dislocations are largely determined by the RFD
configuration, a CDW $2\pi$-dislocation can manifest itself as a
$2\pi$-vortex in either PDW component $\Delta_{\pm Q}$ {[}see Fig.\,\ref{fig:pdw2-cdw}(a)
and (b){]}. This ``selection'' depends on the initial configurations
and the simulated annealing process (for instance, the temperature
intervals during cooling and the MC steps for each temperature). Different
selection would give a slightly different total energy, but it is
understood that in 2D, the interplay of PDW, CDW and RFD always produces
a short range PDW and CDW state with the presence of topological defects.

\section{Phase-separated pair nematic (PSPN) state\label{sec:PDW4}}

In this section, motivated by the STM and REXS experimental results,
we introduce a four-component PDW (PDW4) state with phase separation
in directional components, which is dubbed as a phase-separated pair
nematic (PSPN) state. Experimentally, STM results \cite{Mesaros22072011,Hamidian2015}
reveal that the presence of short range CDW within the pseudogap regime.
These CDW orders are directional and have wave vectors $2Q_{x}$ and
$2Q_{y}$. It is also shown that these two CDW components compete
with each other in real space and form a domain structure, or phase
separation pattern. This picture is also supported by the REXS \cite{Comin20032015}
showing that the results are more consistent with stripy CDW domains.
Viewing this in terms of PDW (section \ref{sec:PDW2}), within each
domain, the directional CDW with $2Q_{a}$ ($a=x,y$) is actually
induced by a PDW2 with components $\Delta_{\pm Q_{a}}$ and this subsystem
can be described by the PDW2 model Eq.\,\eqref{eq:F_2}. However,
for the global system with $\Delta_{x}$ and $\Delta_{y}$ domains
($\Delta_{a}$ denotes $\Delta_{\pm Q_{a}}$ components), we have
to generalize to a PDW4 model \cite{Berg2007,Agterberg2008,Berg2009,Berg2009a,Berg2009b,PhysRevX.4.031017,Agterberg2015,RevModPhys.87.457}.
Moreover, in order to reproduce phase separation pattern, we shall
consider strong competition in the directional components $\Delta_{a}$
{[}$\gamma_{2}^{\Delta}>0$, see Eq.\,\eqref{eq:F4} below{]}.

\begin{figure}
\includegraphics[width=0.95\columnwidth]{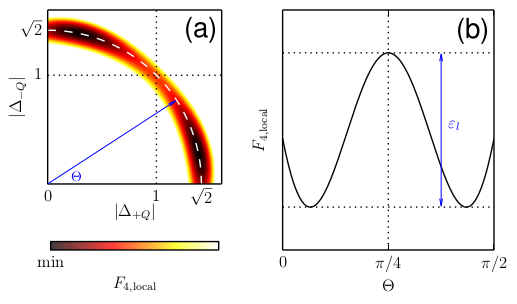}\caption{\label{fig:epsilon_l}(Color online) (a) Schematic density plot of
$F_{4,\mathrm{local}}$ {[}Eq.\,\eqref{eq:F4loc}{]} as a function
of $|\Delta_{\pm Q}|$ for $\gamma_{3}^{\Delta}>h_{\mathrm{rms}}>0$.
(b) Plot of $F_{4,\mathrm{local}}$ as a function of $\Theta$ defined
in (a) along the white dashed line with $\sum_{s}|\Delta_{sQ}|^{2}=-\frac{\alpha^{\Delta}}{2\gamma_{1}^{\Delta}}=2$.
$F_{4,\mathrm{local}}$ has two minima separated by an energy barrier
$\varepsilon_{l}$. At either minimum, the induced CDW $\rho_{2Q}$
and LC order $l$ are both finite.}
\end{figure}

On the other hand, TRSB order is detected in the pseudogap by PKR
\cite{Xia2008,Hashimoto2010} and PND \cite{Fauque2006,Li2008,Mangin-Thro2015}.
The experiments seem to support that the loop current (LC) order \cite{Simon2002}
is responsible for the TRSB order. A natural question would be whether
we can explain this TRSB order within the PDW4 model. Interestingly,
Agterberg \textit{et\,al}.\,\cite{Agterberg2015} proposed that
the LC order can arise from a PDW theory as a secondary composite
order. However, it is readily discovered that within the PDW4 model
the LC order is incompatible with the induced CDW in a PDW state with
spatially homogeneous components $\Delta_{Q}$. One way to resolve
this issue is to introduce an eight-component PDW model, as done by
Agterberg \textit{et\,al}.\,\cite{Agterberg2015}, while another
feasible way, as will be demonstrated below, is to include RFD and
to relax the assumption that the state has homogeneous PDW components. 

In the PDW4 model, another relevant parameter is $\gamma_{3}^{\Delta}$
{[}see Eq.\,\eqref{eq:F4} below{]} controlling whether the $\Delta_{+Q_{a}}$
and $\Delta_{-Q_{a}}$ components compete. If the components $\Delta_{\pm Q_{a}}$
are assumed to be spatially independent, from the known results on
the possible PDW4 ground states \cite{Agterberg2008,Agterberg2015},
we are constrained to choose $\gamma_{3}^{\Delta}<0$ such that $\Delta_{+Q_{a}}$
and $\Delta_{-Q_{a}}$ can coexist. Otherwise, no CDW $\rho_{2Q_{a}}=\Delta_{-Q_{a}}^{*}\Delta_{+Q_{a}}$
can be induced, contradicting the experimental results discussed above.
By introducing disorders in the form of RFD, we break the translational
invariance and allow the PDW components $\Delta_{\pm Q_{a}}$ to be
spatially inhomogeneous. We will demonstrate that in this setting,
the parameter range $\gamma_{3}^{\Delta}>0$ (for competing $\Delta_{+Q_{a}}$
and $\Delta_{-Q_{a}}$) should also be taken into account. Inclusion
of RFD and the parameter choice of $\gamma_{2}^{\Delta},\gamma_{3}^{\Delta}>0$
lead to two important consequences: (i) generating a phase separation
in $\rho_{2Q_{x,y}}$ (induced by $\Delta_{x,y}$) that is consistent
with that observed in the experiments and (ii) resolving the incompatibility
issue of CDW and LC orders within the PDW4 model. Furthermore, we
show that in the PDW4 model there is also a crossover temperature
$T_{CO}<T^{*}$ to account for the short range static CDW within the
pseudogap. For temperature $T_{CO}<T<T^{*}$, the induced CDW is dynamically
fluctuating. The simulation result also suggests that the PDW superconductivity
is absent owing to strong SC phase fluctuations, complying with non-superconducting
pseudogap. These results constitute the properties of the PSPN state. 

We consider the PDW4 system on a square lattice. We shall assume that
the two new components $\Delta_{\pm Q_{x}}$ are related to the PDW2
components $\Delta_{\pm Q}$ with $Q=Q_{y}$ by a $\pi/2$-rotation.
By construction, we require that $Q_{x}\neq Q_{y}$. The order parameter
is given by 
\begin{equation}
\Delta(\mathbf{i})=\sum_{as}\Delta_{sQ_{a}}(\mathbf{i})e^{isQ_{a}\cdot x_{\mathbf{i}}}\thinspace.
\end{equation}
We write down the lattice PDW4 GL functional \cite{RevModPhys.87.457}
by generalizing the PDW2 case,
\begin{eqnarray}
F_{4,\mathrm{lattice}} & = & F_{4,\mathrm{lattice}}^{\Delta}+F_{4,\mathrm{lattice}}^{\rho}\label{eq:F4}
\end{eqnarray}
with
\begin{eqnarray*}
 &  & F_{4,\mathrm{lattice}}^{\Delta}=\\
 &  & -2\kappa_{\parallel}^{\Delta}\sum_{\mathbf{i}as}\text{Re}\left[\Delta_{sQ_{a}}\left(\mathbf{i}+a_{\parallel}\right)\Delta_{sQ_{a}}^{*}\left(\mathbf{i}\right)\right]\\
 &  & -2\kappa_{\perp}^{\Delta}\sum_{\mathbf{i}as}\text{Re}\left[\Delta_{sQ_{a}}\left(\mathbf{i}+a_{\perp}\right)\Delta_{sQ_{a}}^{*}\left(\mathbf{i}\right)\right]\\
 &  & +\tilde{\alpha}^{\Delta}\sum_{\mathbf{i}as}\left|\Delta_{sQ_{a}}\left(\mathbf{i}\right)\right|^{2}\\
 &  & +\gamma_{1}^{\Delta}\sum_{\mathbf{i}}\left(\sum_{as}\left|\Delta_{sQ_{a}}\left(\mathbf{i}\right)\right|^{2}\right)^{2}\\
 &  & +\gamma_{2}^{\Delta}\sum_{\mathbf{i}}\left(\sum_{s}\left|\Delta_{sQ_{x}}\left(\mathbf{i}\right)\right|^{2}\right)\left(\sum_{s}\left|\Delta_{sQ_{y}}\left(\mathbf{i}\right)\right|^{2}\right)\\
 &  & +\gamma_{3}^{\Delta}\sum_{\mathbf{i}a}\left|\Delta_{+Q_{a}}\left(\mathbf{i}\right)\right|^{2}\left|\Delta_{-Q_{a}}\left(\mathbf{i}\right)\right|^{2}\\
 &  & +2\gamma_{4}^{\Delta}\sum_{\mathbf{i}}\mathrm{Re}\left[\Delta_{+Q_{x}}(\mathbf{i})\Delta_{-Q_{x}}(\mathbf{i})\Delta_{+Q_{y}}^{*}(\mathbf{i})\Delta_{-Q_{y}}^{*}(\mathbf{i})\right]
\end{eqnarray*}
\begin{eqnarray*}
 &  & F_{4,\mathrm{lattice}}^{\rho}=\\
 &  & -2h_{\text{rms}}\sum_{\mathbf{i}a}\mathrm{Re}\left[h_{a}(i)e^{-i\eta_{a}(\mathbf{i})}\Delta_{-Q_{a}}^{*}(\mathbf{i})\Delta_{Q_{a}}(\mathbf{i})\right]\\
 &  & -2\kappa_{\parallel}^{\rho}\sum_{\mathbf{i}a}\mathrm{Re}\left[\Delta_{-Q_{a}}\cdotp\Delta_{+Q_{a}}^{*}(\mathbf{i}+a_{\parallel})\Delta_{-Q_{a}}^{*}\cdotp\Delta_{+Q_{a}}(\mathbf{i})\right]\\
 &  & -2\kappa_{\perp}^{\rho}\sum_{\mathbf{i}a}\mathrm{Re}\left[\Delta_{-Q_{a}}\cdotp\Delta_{+Q_{a}}^{*}(\mathbf{i}+a_{\perp})\Delta_{-Q_{a}}^{*}\cdotp\Delta_{+Q_{a}}(\mathbf{i})\right]\\
 &  & +2\left(\kappa_{\parallel}^{\rho}+\kappa_{\perp}^{\rho}\right)\sum_{\mathbf{i}a}\left|\Delta_{-Q_{a}}(\mathbf{i})\Delta_{+Q_{a}}(\mathbf{i})\right|^{2}\thinspace,
\end{eqnarray*}
where we have implicitly replaced $\rho_{2Q_{a}}\rightarrow\Delta_{-Q_{a}}^{*}\Delta_{+Q_{a}}$
and $\Delta_{-Q_{a}}^{*}\cdotp\Delta_{+Q_{a}}(\mathbf{i})$ denotes
$\Delta_{-Q_{a}}^{*}(\mathbf{i})\Delta_{+Q_{a}}(\mathbf{i})$ for
brevity. Here the summation is over $a=x,y$, and $\tilde{\alpha}^{\Delta}=\alpha^{\Delta}+2\kappa_{\parallel}^{\Delta}+2\kappa_{\perp}^{\Delta}$.
The notation $a_{\parallel}=a$ and $a_{\perp}=\hat{y}$ ($\hat{x}$)
if $a=\hat{x}$ ($\hat{y}$). The rest is similar as in the PDW2 case,
except that now we have four PDW components at $\pm Q_{x,y}$ and
allow coupling ($\gamma_{2}^{\Delta}$ term) between the directional
components. Notice that each $2Q_{a}$-CDW is coupled to a different
random field disorder $h_{a}e^{i\eta_{a}}$. 

In the absence of RFD $h_{\mathrm{rms}}=0$, the ground state of the
PDW4 functional is mainly controlled by $\gamma_{2}$ and $\gamma_{3}$.
As in the PDW2 case, the sign of $\gamma_{3}^{\Delta}$ controls whether
two components $\Delta_{+Q_{a}}$ and $\Delta_{-Q_{a}}$ ($a=x,y$)
coexist ($\gamma_{3}^{\Delta}<0$) or compete ($\gamma_{3}^{\Delta}>0$)
with each other. While the sign of $\gamma_{2}^{\Delta}$ dictates
the coexistence ($\gamma_{2}^{\Delta}<0$) or competition ($\gamma_{2}^{\Delta}>0$)
of $\left|\Delta_{x}\right|^{2}$ and $\left|\Delta_{y}\right|^{2}$
components, where $\left|\Delta_{a}\right|^{2}=\sum_{s}\left|\Delta_{sQ_{a}}\right|^{2}$.
In the literature, the main concern is on the latter. (There is also
a similar debate \cite{Robertson2006,DelMaestro2006} on pure CDW
model) In Ref.\,\cite{Berg2009,Berg2009a,Berg2009b}, the authors
propose a PDW theory with $\gamma_{2}^{\Delta}>0$ and $\gamma_{3}^{\Delta}<0$
such that in a clean 2D system, the ground state is a (unidirectional)
striped PDW with either $\Delta_{x}$ or $\Delta_{y}$ component.
On the other hand, Ref.\,\cite{PhysRevX.4.031017} suggests a bidirectional
PDW state with $\gamma_{2}^{\Delta},\gamma_{3}^{\Delta}<0$ that the
$\Delta_{x,y}$ components coexist. Experimentally, X-ray scattering
measurements \cite{Comin20032015,2014arXiv1402.5415C} reveal the
CDW with only wave-vectors $2Q_{x}$ and $2Q_{y}$ but not $\pm2Q_{x}\pm2Q_{y}$.
If the PDW, as well as the induced CDW, is bidirectional, then it
is likely that the $\pm2Q_{x}\pm2Q_{y}$ CDW components would be observed
\cite{Robertson2006} (see, however, Ref.\,\cite{PhysRevX.4.031017}).
Here we propose another parameter choice $\gamma_{2}^{\Delta},\gamma_{3}^{\Delta}>0$
such that all four components $\Delta_{sQ_{a}}$ compete with each
other and constitute the PSPN state. We shall show that it can generate,
besides the CDW orders, a LC order \cite{Simon2002} accounting for
the polar Kerr effect and the intra-unit cell time-reversal symmetry
breaking order observed respectively in the PKR and PND experiments.

The (phenomenological) LC order emerged from PDW order $\Delta_{\pm Q_{a}}$
is defined as \cite{Agterberg2015}
\begin{equation}
l_{a}=\left|\Delta_{+Q_{a}}\right|^{2}-\left|\Delta_{-Q_{a}}\right|^{2}\thinspace.\label{eq:LCO}
\end{equation}
This order parameter is translational invariant (hence giving rise
to an IUC order observed in PND), odd under either time reversal symmetry
and parity, and invariant under their product. (We note that from
a pure CDW order, one can also construct a similar composite TRSB
order \cite{Wang2014,Gradhand2015}) Since this LC order is originated
from the PDW, so its characteristic temperature is the same as $T^{*}$,
which is indeed observed in the PND \cite{Fauque2006,Li2008,Mangin-Thro2015}.
At first sight, in the PDW4 model this order $l_{a}$ is incompatible
with the CDW order $\rho_{2Q_{a}}$. If we restrict ourselves to spatially
independent order, then a finite $l_{a}$ would require $\gamma_{3}^{\Delta}<0$,
which leads to one of the $\Delta_{+Q_{a}}$ and $\Delta_{-Q_{a}}$
components to vanish and eventually the induced CDW order $\rho_{2Q_{a}}=0$.
This is the reason why the authors in Ref.\,\cite{Agterberg2015}
turn to an eight-component PDW theory. For details of the eight-component
theory, the readers shall refer to the reference. Here we will demonstrate
that if we relax the assumption of spatial homogeneity of the order
parameters and introduce RFD, finite orders in both $l_{a}$ and $\rho_{2Q_{a}}$
are actually compatible within the PDW4 model.

Before diving into the numerics, we shall first consider a simplified
model to understand the interplay of (short range) CDW and LC order
with the RFD. The simplified ``on-site'' model is given by
\begin{eqnarray}
F_{4,\mathrm{local}} & = & +\alpha^{\Delta}\sum_{s}\left|\Delta_{sQ}\right|^{2}+\gamma_{1}^{\Delta}\left(\sum_{s}\left|\Delta_{sQ}\right|^{2}\right)^{2}\label{eq:F4loc}\\
 &  & +\gamma_{3}^{\Delta}\left|\Delta_{+Q}\right|^{2}\left|\Delta_{-Q}\right|^{2}-2h_{\mathrm{rms}}\left|\Delta_{-Q}\Delta_{+Q}\right|\nonumber 
\end{eqnarray}
such that we can ignore the phase degrees of freedom. Here we have
already assumed $\gamma_{2}^{\Delta}>0$ and only one pair of $\Delta_{\pm Q}$
exists locally. We further assume that the CDW smectic phase is already
aligned with the local pinning field such that the RFD term reduces
to the present form. Now we notice that although $\gamma_{3}^{\Delta}>0$
would suppress one of the $\Delta_{\pm Q}$ components, but the RFD
term ($h_{\mathrm{rms}}>0$) induces a finite CDW order $\sim\left|\Delta_{-Q}\Delta_{+Q}\right|$
such that both $\Delta_{\pm Q}\neq0$. This competition, provided
with the suitable parameters (see Fig.\,\ref{fig:epsilon_l}), can
result in a local state with both non-vanishing $\rho_{2Q}$ and $l$.
We shall briefly remark that the effect of RFD on the eight-component
PDW theory \cite{Agterberg2015}, in which the incompatibility issue
of CDW and LC order is solved by introducing extra competing PDW components.
If RFD is introduced, then using a similar argument as above, it is
likely that extra CDW components other than those of wave vectors
$2Q_{x}$ and $2Q_{y}$ can be induced. While currently there is no
experimental evidence showing these extra CDW components. On the other
hand, in Fig.\,\ref{fig:epsilon_l} we notice an interesting feature
that the two minima are separated by an energy barrier $\varepsilon_{l}$.
One may define a temperature scale $T_{K}\sim\langle\varepsilon_{l}(\mathbf{i})\rangle_{\mathbf{i}}$
as the spatial average of the energy barriers, which characterizes
the stiffness of the state to (thermally activated) relative amplitude
fluctuations between $|\Delta_{+Q_{a}}$\textbar{} and $|\Delta_{-Q_{a}}$\textbar{}.
For $T<T_{K}$, the energy barrier hinders the relative amplitude
fluctuations and this results in finite LC order. For $T>T_{K}$,
the thermal fluctuations overcome the energy barriers and the system
now consists of fluctuating LC order domains with different signs
leading to a thermal average $\langle\sum_{a}l_{a}\rangle\sim0$ for
TRSB order. This argument might explain the disparity between the
temperature scales for TRSB order measured in PKR and PND, in which
the one measured in PKR is strictly lower than that in PND (see Introduction).
The elaborated study however will be pursued elsewhere.

Now back to the entire system described by $F_{4,\mathrm{lattice}}$
{[}Eq.\,\ref{eq:F4}{]} with RFD $h_{\mathrm{rms}}h(R)e^{i\eta(R)}$,
we focus on those sites with particularly strong pinning strengths.
At one of these sites, to gain the pinning energy, the CDW smectic
phase would align with the RFD phase $\eta$ (see section \ref{sub:PDW2-with-RFD}).
The PDW and CDW stiffness terms would induce a finite region of approximately
the same smectic phase around the strong disorder site. Now if we
are restricted to the parameter choice discussed above and take the
directional PDW into account, we would then expect a phase separated
domain pattern, in which the directional PDW, together with the induced
directional CDW and LC orders, presents due to the ``nucleation''
around these strong disorders. The phase separated orders discussed
above can be regarded as imposing phase separation patterns on the
PDW2 orders. This indicates that the correlations of the PDW and CDW
orders are bounded by those in PDW2, and thus the resultant state
also has PDW and CDW with the short range exponential correlations.
Analog to the PDW2 case, we also expect a crossover temperature for
the short range static CDW in the state. Owing to the phase separation
pattern and the local nematicity originated from PDW superconductivity,
we shall call this the phase-separated pair nematic (PSPN) state.
Below we will demonstrate that the PSPN state can be resulted from
the PDW4 GL functional with RFD {[}Eq.\,\ref{eq:F4}{]} via MC simulation
study.

\subsection{Numerics: Monte Carlo study of PSPN\label{sub:PSPN-numerics}}

\begin{figure*}
\begin{centering}
\includegraphics[width=0.6\textwidth]{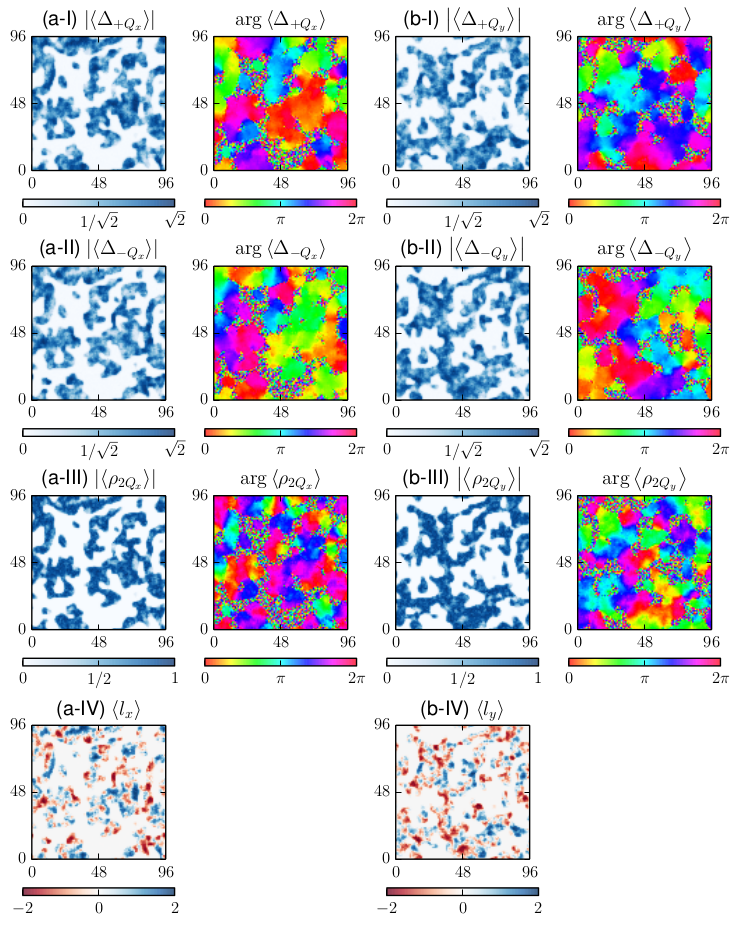}
\par\end{centering}

\caption{\label{fig:pdw4}(Color online) Monte Carlo results of the PSPN state
in the presence of RFD for the averaged order parameters in (a) $x$-
and (b) $y$-directions, using GL functional $F_{4,\mathrm{lattice}}$
Eq.\,\eqref{eq:F4}. We plot the spatial dependent (I and II) PDW
components $\left\langle \Delta_{\pm Q_{a}}\right\rangle $, (III)
CDW $\left\langle \rho_{2Q_{a}}\right\rangle $ and (IV) LC order
$\left\langle l_{a}\right\rangle $. The PDW and CDW orders are phase
separated in the directional components. Also the PDW and CDW phases
$\arg\langle\Delta_{sQ_{a}}\rangle$ and $\arg\langle\rho_{2Q_{a}}\rangle$
form a domain pattern, similar to the PDW2 case. Moreover, finite
CDW and LC orders are induced in the system. In order to capture the
effects of $\gamma_{2,3}^{\Delta}>0$, we allow updates on both amplitudes
and phases of the PDW order parameters $\Delta_{sQ_{a}}$. Parameters
(in meV): $\alpha^{\Delta}=-100$, $\gamma_{1}^{\Delta}=25$, $\gamma_{2}^{\Delta}=20$,
$\gamma_{3}^{\Delta}=5$, $\gamma_{4}^{\Delta}=0$, $\kappa_{\parallel,\perp}^{\Delta}=10$,
$\kappa_{\parallel,\perp}^{\rho}=1$, $h_{\mathrm{rms}}=2$. System
size: $96\times96$ with periodic boundary condition. Simulated annealing
is performed from a high temperature to the final temperature at $10$\,K.}
\end{figure*}

\begin{figure}
\begin{centering}
\includegraphics[width=0.65\columnwidth]{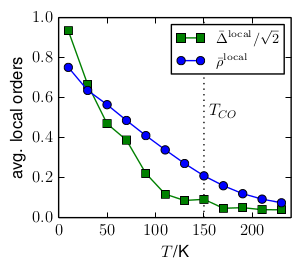}
\par\end{centering}

\caption{(Color online) \label{fig:local-orders} Melting of short range PDW
and CDW in the PSPN state in the presence of RFD by performing MC
simulations on GL functional $F_{4,\mathrm{lattice}}$ {[}Eq.\,\eqref{eq:F4}{]}.
As a function of temperature, the averaged \textit{local} PDW and
CDW orders, defined as $\bar{\Delta}^{\mathrm{local}}=\frac{1}{N}\sum_{\mathbf{i}}\sqrt{\sum_{sa}|\Delta_{sQ_{a}}(\mathbf{i})|^{2}}$
and $\bar{\rho}^{\mathrm{local}}=\frac{1}{N}\sum_{\mathbf{i}a}|\rho_{2Q_{a}}(\mathbf{i})|$,
are plotted, where $N$ is total number of sites. Notice that the
plotted values are rescaled by their maximal possible values. In order
to study the effect of the phase fluctuations, the amplitudes of the
PDW components $|\Delta_{sQ_{a}}|$ are taken from the averaged results
in Fig.\,\ref{fig:pdw4} as input and kept constant during the simulations,
while only the phases $\arg\Delta_{sQ_{a}}$ are updated. We set the
threshold $\rho_{0}=0.2$ \cite{Note1} for the local CDW order to
define the crossover temperature $T_{CO}$ for the short range static
CDW orders. Parameters are the same as in Fig.\,\ref{fig:pdw4}.}
\end{figure}

To better illustrate our points, we perform the MC simulations similar
to the PDW2 case. We consider the case of $\gamma_{2,3}^{\Delta}>0$
as we discussed above and choose $\alpha^{\Delta}$ and $\gamma_{1}^{\Delta}$
such that the average $\left|\Delta_{sQ_{a}}\right|\sim1$. We divide
the MC study into two stages. In the first stage, in order to capture
the effects of $\gamma_{2,3}^{\Delta}$ on the resultant state, we
allow updates (Metropolis algorithm, single site) on both the amplitudes
and phases of the PDW order parameters. We start with a random phase
configuration with $\left|\Delta_{sQ_{a}}\right|=1$ and perform the
simulated annealing process to let the system to better equilibrate
at the final temperature. For later comparison to experimental data,
the simulation is performed with a typical RFD configuration. We then
obtain a state at low temperature, resembling the ground state for
a given set of parameters. In the second stage, we are interested
in the effect of thermal fluctuations on the orders, in particular
the phase fluctuations induced by proliferation of topological defects
(see section \ref{sub:Topo-defects}). Analog to the studies of BKT
transition \cite{Berezinskii1971,KosterlitzJ.M.;Thouless1973} (see
also random field XY model \cite{Gingras1996,PhysRevX.5.031008}),
we shall assume that the system is deep in the state obtained in stage
one. We thus utilize the averaged amplitudes as input (i.e.\,the
amplitudes are kept fixed during the MC) and only perform MC simulations
with phase updates (Metropolis algorithm, single site) for various
temperatures. To be consistent, $\alpha^{\Delta}$ and $\gamma_{1}^{\Delta}$
used in stage one should be chosen to suppress the the total PDW amplitude
fluctuations but still permit sufficient acceptance rates for sampling
states with various relative PDW amplitudes. Besides, directional
CDW with wave vectors $2Q_{x,y}$ requires that $\gamma_{2}^{\Delta}$
is dominated over $\gamma_{3}^{\Delta}$, while in order to induce
finite orders in both CDW and LC order, $\gamma_{3}^{\Delta}$ and
$h_{\mathrm{rms}}$ should be comparable (see Fig.\,\ref{fig:epsilon_l}).
We shall set $\gamma_{4}^{\Delta}=0$ such that the phases of the
directional PDW are not coupled \footnote{More precisely, $\gamma_{4}^{\Delta}$ controls the phase coherence
between the induced directional charge-4e SC $\Delta_{4e}^{a}=\Delta_{+Q_{a}}\Delta_{-Q_{a}}$.
Choosing $\gamma_{4}^{\Delta}=0$ destroys the phase coherence and
results in a phase separation in the directional charge-4e SC. More
detailed discussion is given in section I of SM \cite{Note2}.}.

In Fig.\,\ref{fig:pdw4}, we show the simulation results in stage
one for a typical RFD configuration for PDW4. Several observations
are followed. Owing to a large $\gamma_{2}^{\Delta}>0$, the two components
$\Delta_{x,y}$ compete with each other and it results in phase separation
for $\Delta_{x,y}$, as well as $\rho_{2Q_{x,y}}$. (Recall that $\Delta_{a}$
denotes directional PDW components $\Delta_{\pm Q_{a}}$) Within each
domain of $\Delta_{x,y}$ and $\rho_{2Q_{x,y}}$, the phases ($\arg\Delta_{\pm Q_{a}}$
and $\arg\rho_{2Q_{a}}$) have similar domain structures as in the
PDW2 case, except that there are regions of vanishing amplitude in
which the phases are not well-defined. There are still topological
defects in the domains, but the number is not well constrained due
to the regions of vanishing amplitude. Experimentally, CDW dislocations
of $2\pi$ windings are indeed observed in STM \cite{Mesaros22072011}
and they could trap the half SC fluxes for a PDW state (see section
\ref{sub:Topo-defects}), provided that the SC phase is strongly pinned.
One can probe these half fluxes by SQUID microscopy and test the proposed
PDW theory. Next, we notice both $\rho_{2Q_{a}}$ and $l_{a}$ have
finite magnitude locally, in accord with our discussion on the simplified
model Eq.\,\eqref{eq:F4loc}. Also, as discussed, the PDW and CDW
orders are expected to be short ranged and decay exponentially \cite{Gingras1996,PhysRevX.5.031008}
(see Fig.\,1 in Supplementary Material \footnote{See Supplementary Material at {[}\textit{URL}{]} for numerically computed
correlations in directional CDW and charge-4e SC, comparisons with
experimental data and more details on the superposition feature of
the spectral functions.} (SM) for numerical evidence and more discussions), due to the random
phases and presence of topological defects. The LC order $l_{a}$
are inhomogeneous and have random signs. This selection of the local
sign is related to the initial configuration and simulated annealing
process. We also note that the amplitudes $\left|\Delta_{sQ_{a}}\right|$
are inhomogeneous, to accommodate both non-vanishing $\rho_{2Q_{a}}$
and $l_{a}$. Again, the simulation might be trapped at some local
energy minimum and unable to attain the true ground state, but the
main purpose is to illustrate the effect of RFD and the simultaneous
presence of $l_{a}$ and $\rho_{2Q_{a}}$ for our choice of $\gamma_{3}^{\Delta}>0$.

\begin{figure}
\begin{centering}
\includegraphics[width=1\columnwidth]{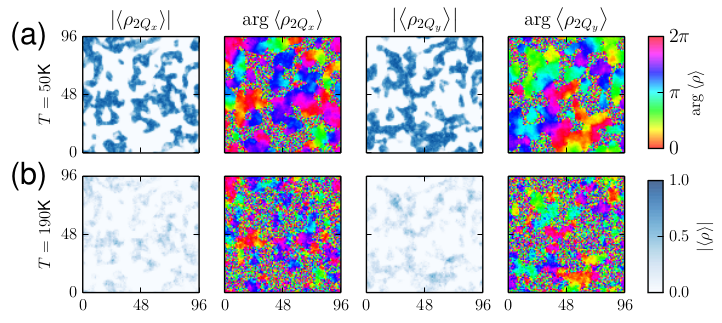}
\par\end{centering}

\caption{\label{fig:pdw4-Tco}(Color online) Spatial variations of CDW in the
PSPN state at two temperatures: (a) $T=50\thinspace\mathrm{K}<T_{CO}$
and (b) $T=190\thinspace\mathrm{K}>T_{CO}$. In the same MC simulations
of Fig.\,\ref{fig:local-orders} using GL functional $F_{4,\mathrm{lattice}}$
{[}Eq.\,\eqref{eq:F4}{]}, we show the results for the averaged directional
CDW $\langle\rho_{2Q_{a}}\rangle$. The noticeable difference for
these two temperatures is the vanishing small magnitudes of the averaged
CDW at $T=190\thinspace\mathrm{K}$ due to the strong thermal fluctuations
in the smectic phases for $T>T_{CO}$.}
\end{figure}

\begin{figure}
\begin{centering}
\includegraphics[width=1\columnwidth]{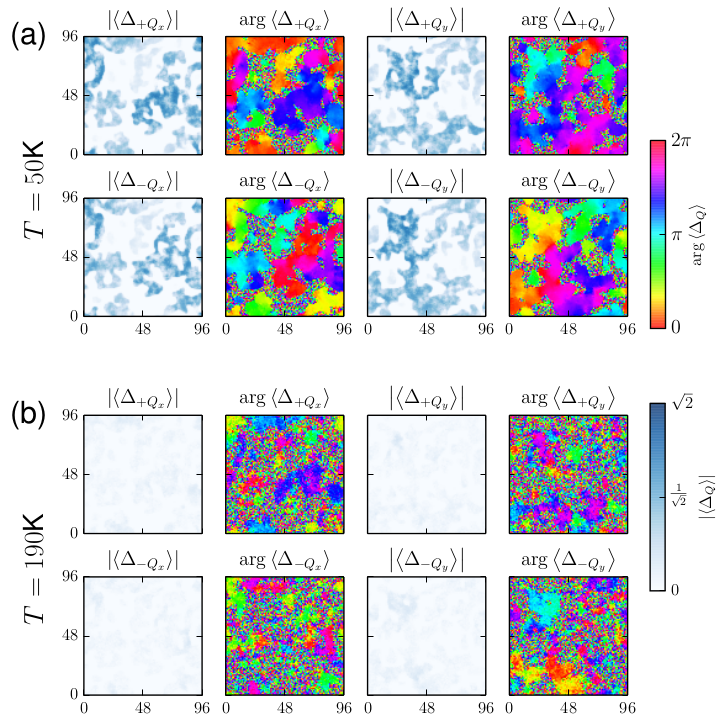}
\par\end{centering}

\caption{\label{fig:pdw4-pdwSC}(Color online) Spatial variations of PDW components
in the PSPN state at two temperatures: (a) $T=50\thinspace\mathrm{K}<T_{CO}$
and (b) $T=190\thinspace\mathrm{K}>T_{CO}$. In the same MC simulations
in Fig.\,\ref{fig:local-orders} using GL functional $F_{4,\mathrm{lattice}}$
{[}Eq.\,\eqref{eq:F4}{]}, we measured the PDW order parameters $\left\langle \Delta_{\pm Q_{x,y}}\right\rangle $.
At $T=50\thinspace\mathrm{K}$, some domains of directional PDW are
wiped out owing to strong fluctuations in SC phases. At $T=190\thinspace\mathrm{K}$,
the fluctuations are stronger and most of the PDW orders are destroyed.
These results suggest the absence of PDW superconductivity irrespective
of $T_{CO}$.}
\end{figure}

We then move on to stage two to study the effect of the thermal phase
fluctuations on the PDW and CDW orders. As discussed, owing to the
pinning of the RFD, the PDW and CDW orders are short ranged, and we
expect a crossover temperature $T_{CO}$ for the static CDW. To determine
$T_{CO}$, we plot the temperature dependent \textit{local} CDW order
$\bar{\rho}^{\mathrm{local}}$, as well as local order $\bar{\Delta}^{\mathrm{local}}$
for PDW, in Fig.\,\ref{fig:local-orders}. The local orders decay
in a continuous fashion as temperature increases. At low temperature,
the local charge orders with significant amplitude are still present,
as shown in Fig.\,\ref{fig:pdw4-Tco}(a), due to the strong pinning
effect of the RFD and correlation effect of the stiffness terms $\kappa^{\Delta,\rho}$.
At high enough temperature, the local charge orders are almost absent
{[}Fig.\,\ref{fig:pdw4-Tco}(b){]}, as the thermal phase fluctuations
have overcome the pinning effect from RFD and stiffness terms. We
can view this as the thermally activated diffusion of the topological
defects (dislocations) of the CDW. Moreover, although the amplitudes
are kept fixed as the averaged values taken from Fig.\,\ref{fig:pdw4},
we notice that as a manifestation of the strong phase fluctuations,
the domains of well-defined CDW smectic phase at high temperature
{[}Fig.\,\ref{fig:pdw4-Tco}(b){]} shrink considerably. We shall
roughly identify the crossover temperature for the short range static
CDW $T_{CO}=150\thinspace\mathrm{K}$ by setting the threshold $\rho_{0}=0.2$
\footnote{This threshold is set in analog of the MC study of the BKT transition,
in which a \textit{global} order parameter \unexpanded{$\bar{\Delta}=\langle|N^{-1}\sum_{\mathbf{i}}e^{i\theta(\mathbf{i})}|\rangle$}
is used ($\theta$ is the XY phase and $N$ is number of sites). Near
the BKT transition temperature, there is a sharp drop in $\bar{\Delta}$
from $0.6$ to $0.2$. Thus for $\bar{\Delta}\gtrsim0.2$, the system
is regarded as in the short-range correlated phase.}. In this sense, Fig.\,\ref{fig:pdw4-Tco}(a) and (b) actually correspond
to the spatial variations of the directional CDW for $T<T_{CO}$ and
$T>T_{CO}$ respectively.

Next we discuss whether the PDW superconductivity can survive the
thermal phase fluctuations. Fig.\,\ref{fig:pdw4-pdwSC} shows the
MC results for the magnitudes and phases of the averaged PDW orders
$\langle\Delta_{sQ_{a}}\rangle$ at $T=50\thinspace\mathrm{K}$ and
$190\thinspace\mathrm{K}$. At $T=50\thinspace\mathrm{K}<T_{CO}$
{[}Fig.\,\ref{fig:pdw4-pdwSC}(a){]}, there are PDW domains with
almost vanishing amplitudes, in contrast to the same area in Fig.\,\ref{fig:pdw4-Tco}(a),
in which charge orders of significant amplitude still remain. This
difference is because of the strong pinning on the CDW smectic phase
by the RFD, while PDW superconducting phases $\vartheta_{a}\equiv\frac{1}{2}(\theta_{+Q_{a}}+\theta_{-Q_{a}})$
are not (here $\theta_{sQ_{a}}\equiv\arg\Delta_{sQ_{a}}$ are the
phase fields of the PDW components, see also section \ref{sub:Topo-defects}).
Since the PDW smectic phases $\varphi_{a}\equiv\frac{1}{2}(\theta_{+Q_{a}}-\theta_{-Q_{a}})$
are also pinned indirectly by the RFD through the induced CDW below
$T_{CO}$, this suggests that the fluctuations in the SC phases $\vartheta_{a}$
in the domains are strong. We note that the phase fluctuations are
stronger than the PDW2 case due to the phase separation pattern \cite{Note1}.
The thermal fluctuations can ``invade'' the orders at the boundaries
and destroy them more severe than the system without phase separation.
At $T=190\thinspace\mathrm{K}>T_{CO}$ {[}Fig.\,\ref{fig:pdw4-pdwSC}(b){]},
the thermal fluctuations are severe and the PDW local order is mostly
wiped out, as evident from the vanishing averaged PDW order parameters
and the considerable shrinkage in the PDW domains of well-defined
phases. It is well known that dynamic vortices in a superconductor
dissipate supercurrent, leading to a finite resistance (known as the
\textit{flux flow resistance}, see e.g.\,Ref.\,\cite{Kim1965}).
In this regard, the results in Fig.\,\ref{fig:pdw4-pdwSC} hint that
the PDW superconductivity is absent owing to strong thermal fluctuations
in the superconducting phase irrespective of the crossover temperature
$T_{CO}$ and appear to be consistent with the known fact that the
pseudogap is non-superconducting.

We conclude this section by comparing these results with experiments.
First, RIXS \cite{Chang2012,PhysRevLett.109.167001,PhysRevLett.110.187001}
and STM \cite{Parker2010} reveal traces of the crossover temperature
$T_{CO}$ for static charge order inside the pseudogap. RIXS uses
hard X-rays to measure the static order of a given structure. Similarly
STM measures the static electronic orders. $T_{CO}$'s found in these
probes show a similar trend that is peaked at some doping. Comparing
with our results, we identify the order to be the short range CDW.
REXS measurements however use soft X-rays, that is sensitive to fluctuating
orders, and it can detect the fluctuating CDW up to a higher temperature
$T^{*}$, at which the PDW orders vanish. Next, a smectic modulation
measure $O_{s}^{Q}(r,e)$ in STM \cite{Lawler2010,Mesaros22072011,Hamidian2015}
signifies relative strength of the local smectic (directional) electronic
orders. As discussed at the section overview, these results point
to a phase-separated directional PDW state. We found that using our
model of PDW4 and RFD together with suitable parameters, we can reproduce
the phase separation pattern for the directional CDW (see Fig.\,3
in SM \cite{Note2} for more details). We note that REXS result \cite{Comin20032015}
also supports such a phase separation picture for the directional
CDW.

\section{Constraint from CDW with predominantly d-wave form factor\label{sec:Constraint}}

Before the discussion of the spectral functions of the PDW system,
we shall first figure out the pairing symmetry of the PDW. Both STM
\cite{Fujita2014,Hamidian2015} and REXS \cite{2014arXiv1402.5415C}
experiments strongly indicate that the CDW order within the pseudogap
has a predominantly $d$-wave form factor. If we insist that the CDW
order is only originated from the corresponding PDW order, then we
need to understand how this CDW form factor would constrain the pairing
symmetry of the PDW.

Here we propose that the PDW has a $s'\pm id$-symmetry. First we
write down the definition of a generic SC order parameter at sites
$r_{i}$ and $r_{j}$
\begin{eqnarray}
\Delta\left(r_{1},r_{2}\right) & = & \sum_{Q}\left(\sum_{k}\Delta_{Q}\left(k\right)e^{ik\left(r_{1}-r_{2}\right)}\right)e^{iQ\frac{r_{1}+r_{2}}{2}}\nonumber \\
 & \equiv & \sum_{Q}\Delta_{Q}\left(r\right)e^{iQ\cdot R}\thinspace,\label{eq:Delta_def}
\end{eqnarray}
in which we allow multiple components with different modulation wave
vectors $Q$, which determine the modulation in the average coordinate
$R=\frac{1}{2}\left(r_{1}+r_{2}\right)$, and their (orbital) pairing
symmetries involving relative coordinate $r=r_{1}-r_{2}$ are described
by the form factor $\Delta_{Q}(k)$. We can similarly define the order
parameter for charge modulations \cite{Fujita2014} by a simple replacement
$\Delta\rightarrow\rho$. We note that the definition above only describe
the order parameters of the PDW components $\Delta_{Q}$ spatially
uniform in the average coordinate (i.e.\,$\Delta_{Q}$ is independent
of $R$). If we want to include (slow) spatial variation in $\Delta_{Q}$,
we can add a spatial varying factor by the substitution $\Delta_{Q}(r)\rightarrow\Delta_{Q}(R)f_{Q}(r)$
with the form factor encoded in $f_{Q}(r)$. In this section, we shall
consider spatially uniform case only.

The pairing symmetry factor for $+Q_{x}$ is assumed to be 
\begin{equation}
\Delta_{+Q_{x}}(k)=\Delta_{s}(k_{x}^{2}+k_{y}^{2})+\Delta_{d}e^{i\phi}(k_{x}^{2}-k_{y}^{2})\thinspace,\label{eq:pair_sym_cont}
\end{equation}
where $\Delta_{s,d}\in\mathbb{R}$ , $\phi\in[0,2\pi)$. It is clear
that the first (second) term is $s$-wave ($d$-wave). Here $\Delta_{s,d}$
and $\phi$ will be constrained by the CDW dominating $d$-wave form
factor shown in the experiments. Time reversal symmetry requires that
$\Delta_{-Q}(k)=\Delta_{+Q}^{*}(-k)$. By the relation $\rho_{2Q_{a}}(r)=\Delta_{-Q_{a}}^{*}(r)\Delta_{+Q_{a}}(r)$,
the CDW form factor can be shown as (see Appendix \ref{sec:CDW-form-factor})
\begin{eqnarray}
\rho_{2Q_{x}}(k) & = & (\Delta_{s}^{2}+e^{2i\phi}\Delta_{d}^{2}/2)\underset{s\text{-wave}}{\underbrace{(k_{x}^{2}+k_{y}^{2})^{2}}}\nonumber \\
 &  & +(2e^{i\phi}\Delta_{s}\Delta_{d})\underset{d\text{-wave}}{\underbrace{(k_{x}^{2}+k_{y}^{2})(k_{x}^{2}-k_{y}^{2})}}\nonumber \\
 &  & +(e^{2i\phi}\Delta_{d}^{2}/2)\underset{g\text{-wave}}{\underbrace{(k_{x}^{4}-6k_{x}^{2}k_{y}^{2}+k_{y}^{4})}}
\end{eqnarray}
up to an overall factor. The STM experiment \cite{Fujita2014,Hamidian2015}
reveals that the $s$-wave component is much smaller than the $d$-wave
one, while the REXS experiment \cite{2014arXiv1402.5415C} shows that
a pure $d$-wave CDW cannot fully reproduce the experimental data.
Inspired by these experimental facts, we simply choose the $s$-wave
component $\Delta_{s}^{2}+\frac{1}{2}e^{2i\phi}\Delta_{d}^{2}$ to
vanish and leaving only the $d$-wave and $g$-wave components. This
particular choice requires $e^{i\phi}=\pm i$ and $\Delta_{d}/\Delta_{s}=\sqrt{2}$.
Explicitly we take 
\begin{equation}
\Delta_{sQ_{a}}(k)=\Delta_{0}\left[(k_{x}^{2}+k_{y}^{2})+is\sqrt{2}(k_{x}^{2}-k_{y}^{2})\right]\thinspace,
\end{equation}
where $\Delta_{0}\in\mathbb{C}$. Here $\Delta_{+Q_{a}}$ and $\Delta_{-Q_{a}}$
transform to each other under time reversal symmetry and we simply
choose $\Delta_{sQ_{x}}\leftrightarrow\Delta_{sQ_{y}}$ related by
a $\pi/2$-rotation. In the experiments, the $d$-wave form factor
also indicates that the charge orders are located at the oxygen site.
Therefore, in the (single band) model on a square lattice, we shall
consider the (nearest neighbor) bond order between the lattice sites
(with Cu atoms) to mimic this experimental fact, and we restrict ourselves
to the $s'\pm id$-wave pairing symmetry
\begin{equation}
\frac{\Delta_{sQ_{a}}(k)}{\Delta_{0}}=(\cos k_{x}+\cos k_{y})+is\sqrt{2}(\cos k_{x}-\cos k_{y})\thinspace.\label{eq:pair_sym}
\end{equation}
A detailed derivation of the induced CDW form factor for Eq.\,\eqref{eq:pair_sym}
is given in the Appendix \ref{sec:CDW-form-factor}, in which we show
that the above consideration remains valid in the lattice scenario.

\section{Quasiparticle Spectral Functions and ARPES\label{sec:ARPES}}

In this section, we study the spectral functions of PDW systems and
compare the results with the ARPES experiments. Many exotic properties
are observed in ARPES. Early studies have revealed that the Fermi
surface in the pseudogap forms the so-called Fermi arcs \cite{Norman1998,Comin24012014},
disconnected segments of gapless quasiparticle excitations near the
nodes. Near the antinodes, the spectrum is gapped, while the quasiparticle
peak is very broad or even ill-defined \cite{Orenstein21042000}.
Recently, it is found \cite{Hashimoto2010,He25032011} that by comparing
the data above and below $T^{*}$, the quasiparticle spectra have
the feature of $k_{F}$-$k_{G}$ misalignment near the antinodes,
where $k_{F}$ is the Fermi momentum for $T>T^{*}$ and $k_{G}$ is
the back-bending momentum for $T<T^{*}$. This feature cannot be explained
by conventional SC pairing, otherwise we would expect $k_{F}=k_{G}$.
Moreover, away from the antinode, the spectral gap is closed from
below \cite{Yang2008,PhysRevLett.107.047003,He25032011}. While from
a calculation \cite{PhysRevX.4.031017} for a model with pure CDW,
an order observed in STM and REXS, shows that the spectral gap is
closed from above and it fails to capture this ARPES feature. On the
other hand, PDW models \cite{PhysRevX.4.031017,Agterberg2015} can
successfully capture the $k_{F}$-$k_{G}$ misalignment and the feature
of the gap closing from below, suggesting that PDW plays an important
part in the pseudogap phenomenology.

To compute the spectral functions, we introduce a lattice hamiltonian
with PDW pairings {[}Eq.\,\eqref{eq:ham}{]} taking into account
the spatial variations in the PDW order parameters, pairing symmetry
and thermal (phase) fluctuations. In order to study separately the
effects of the RFD induced random phases, phase separation, and the
thermal fluctuations, we will consider four PDW systems: (i) uniform
PDW2 with $s^{\prime}\pm id$-wave pairing symmetry, (ii) $s^{\prime}\pm id$-wave
PDW2 with random phases induced by RFD, (iii) $s'\pm id$-wave PDW4
with phase separation in pairing amplitudes $\Delta_{x,y}$ but no
random phase, and lastly (iv) PSPN state with thermal fluctuations
in the PDW phases $\theta_{sQ_{a}}$, originated from a $s^{\prime}\pm id$-wave
PDW4 state. For the former three systems, it is sufficient to compute
the spectral functions by feeding a static PDW configuration into
the hamiltonian {[}Eq.\,\eqref{eq:SpecFun}{]}. But for last one,
it is needed to incorporate disorder average and thermal fluctuations
through a sequence of MC generated PDW configurations {[}Eq.\,\eqref{eq:MC-SpecFun}{]}.

Actually there are similar studies on the effect of thermal fluctuations
on the spectral features. In Ref.\,\cite{Bieri2009}, the authors
studied the thermal averaged spectral functions for cuprates within
a spin liquid scenario, where the order parameter fluctuates between
a $d$-wave superconductor and a nearby staggered-flux state related
by $\mathrm{SU}(2)$ rotations. While in Ref.\,\cite{Han2010}, the
authors studied the fluctuating $d$-wave superconductor scenario,
in which the SC phases are strongly disordered due to vortex proliferation.
These studies also appear to reproduce the Fermi arc feature observed
in ARPES.

We first write down the real space hamiltonian $H=H_{0}+H_{\Delta}$
with 
\begin{eqnarray}
H_{0} & = & -\sum_{\mathbf{i}\delta\sigma}t_{\delta}c_{\mathbf{i}+\delta\sigma}^{\dagger}c_{\mathbf{i}\sigma}+\mathrm{h.c.}-\mu\sum_{\mathbf{i}\sigma}c_{\mathbf{i}\sigma}^{\dagger}c_{\mathbf{i}\sigma}\label{eq:ham}\\
H_{\Delta} & = & \sum_{ia}\Delta^{*}(\mathbf{i}+a,\mathbf{i})\left(c_{\mathbf{i}+a\uparrow}c_{\mathbf{i}\downarrow}-c_{\mathbf{i}+a\downarrow}c_{\mathbf{i}\uparrow}\right)+\mathrm{h.c.}\thinspace,\nonumber 
\end{eqnarray}
where $\Delta\left(\mathbf{i}+a,\mathbf{i}\right)$ is the nearest
neighbor singlet (bond) pairing with $a=\hat{x},\hat{y}$. The parameters
are taken from Ref.\,\cite{Comin24012014} that (with lattice constant
$a_{0}=1$) the nearest neighbor hopping $t_{1}=0.4$\,eV, the next
nearest neighbor hopping $t_{2}/t_{1}=-0.2$ and the third nearest
neighbor hopping $t_{3}/t_{1}=0.05$. The PDW order parameters in
momentum space $\Delta_{\pm Q_{a}}(k)$ describe pairing with total
momentum $\pm Q_{a}$ and the pairing centers are at $\pm Q_{a}/2$.
In order to explain the ARPES results, we follow Ref.\,\cite{PhysRevX.4.031017}
and choose the pairing centers at the Brillouin zone (BZ) boundary
that $\pm Q_{x}/2=\pm\left(\pi,Q_{0}/2\right)$ and $\pm Q_{y}/2=\mp\left(Q_{0}/2,\pi\right)$
related by a $\pi/2$-rotation (see Fig.\,\ref{fig:BZ}). In this
study, we take the chemical potential $\mu/t_{1}\approx-0.758$ (corresponding
hole doping $\approx0.117$) such that the electronic band intersects
with the BZ boundary exactly at the aforementioned pairing centers
with $Q_{0}=\pi/4$. The resultant PDW has a period of $8a_{0}$ and
corresponding CDW with period $4a_{0}$. As (hole) doping increases,
the corresponding CDW modulation $2Q_{0}$ decreases by choice \cite{PhysRevX.4.031017}
and this trend is in accord with the results observed in X-ray experiments
\cite{PhysRevLett.110.137004,Comin24012014,DaSilvaNeto2014}.

In order to incorporate spatial non-uniform pairing order parameters,
we generalize Eq.\,\eqref{eq:Delta_def} to 
\begin{equation}
\Delta(r_{1},r_{2})=\bar{\Delta}\sum_{Q}\Delta_{Q}(R)f_{Q}(r)e^{iQ\cdot R}\thinspace,\label{eq:PDWorder}
\end{equation}
where $R=\frac{1}{2}(r_{1}+r_{2})$ is the average coordinate, $r=r_{1}-r_{2}$
is the relative coordinate, $f_{Q}(r)$ describes the pairing symmetry
in real space, given by {[}Eq.\,\eqref{eq:pair_sym}{]}
\begin{equation}
f_{sQ_{a}}(r)=\begin{cases}
1+is\sqrt{2}\thinspace, & r=\pm\hat{x}\\
1-is\sqrt{2}\thinspace, & r=\pm\hat{y}
\end{cases}\thinspace.
\end{equation}
Here the spatial dependent pairing order parameters $\Delta_{Q}(R)\in\mathbb{C}$
will be taken from the results of the MC simulations with spatial
mean magnitude $\langle|\Delta_{Q}(R)|\rangle_{R}=1$. We shall assume
that the orders are sufficiently smooth and we can take the interpolation
value $\Delta_{Q}(R)=\frac{1}{2}\left(\Delta_{Q}(r_{1})+\Delta_{Q}(r_{2})\right)$
as the bond order parameters \footnote{We note that the order parameters of the PDW components $\Delta_{Q}(R)$
in the hamiltonian Eq.\,\eqref{eq:ham} actually take values at the
bond center $R=\frac{1}{2}(r_{1}+r_{2})$ {[}see Eq.\,\eqref{eq:PDWorder}{]},
where $r_{1}$ and $r_{2}$ are adjacent sites. However, the order
parameters $\Delta_{Q}(r_{i})$ given by the MC results are only defined
at the sites $r_{i}$, not at the bond center $R$. In order to connect
the MC data to the hamiltonian, we interpolate $\Delta_{Q}(r_{i})$
and take $\Delta_{Q}(R)=\frac{1}{2}(\Delta_{Q}(r_{1})+\Delta_{Q}(r_{2}))$
at each bond $(r_{1},r_{2})$. This interpolation process should be
justified by the spatial smoothness of $\Delta_{Q}(r_{i})$, at least
within the PDW domains.}. Finally, an overall scaling $\bar{\Delta}$ is added to control
the ``gap'' size. In the following, we compute the quasiparticle
spectral functions in momentum space corresponding to the hamiltonian
Eq.\,\eqref{eq:ham} according to 
\begin{eqnarray}
A_{\left\{ \Delta\right\} }\left(k,\thinspace\omega\right) & = & -\frac{1}{\pi}\text{Im}\left\langle k\middle|\frac{1}{\omega+i0^{+}-H_{\left\{ \Delta\right\} }}\middle|k\right\rangle \thinspace,\label{eq:SpecFun}
\end{eqnarray}
provided that the pairing order parameters $\left\{ \Delta\right\} =\left\{ \Delta_{sQ_{a}}\right\} $
are given. Here $\left|k\right\rangle =\frac{1}{\sqrt{N}}\!\sum_{\mathbf{i}}e^{ik\cdot x_{\mathbf{i}}}\!\left|x_{\mathbf{i}}\right\rangle $
is the single-particle wave vector with momentum $k$ in the first
BZ. 

\begin{figure*}
\begin{centering}
\includegraphics[width=0.85\textwidth]{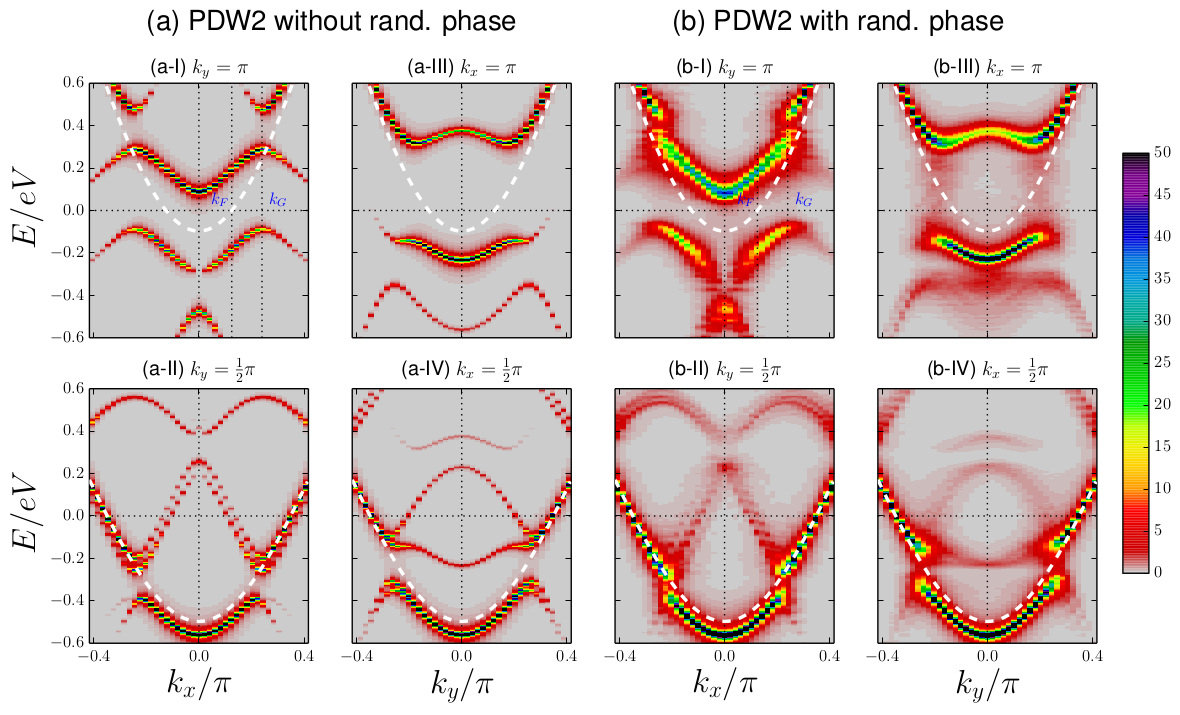}
\par\end{centering}

\caption{\label{fig:PDW2-ARPES}(Color online) Spectral functions at two line
scans (see Fig.\,\ref{fig:BZ}) in BZ. Panel (a) shows the PDW2 system
without adding random phases and panel (b) shows the PDW2 with static
random phases taken from Fig.\,\ref{fig:pdw2-cdw}. The PDW orders
$\Delta_{\pm Q_{y}}$ are added at two pairing centers $\pm Q_{y}/2$.
The white dash lines represent the bare bands. Both systems show the
$k_{F}$-$k_{G}$ misalignment in the $k_{y}=\pi$ plots as seen in
subplots (a-I) and (b-I), where $k_{F}$ is the Fermi momentum of
the bare band and $k_{G}$ is the back bending momentum. But the spectra
in the system with random phases are more broadened. We take $Q_{0}=\pi/4$,
$\bar{\Delta}=0.125\thinspace$eV, $t_{1}=0.4\thinspace\mathrm{eV}$,
$t_{2}/t_{1}=-0.2$, $t_{3}/t_{1}=0.05$, and $\mu/t_{1}\approx-0.758$.
System size: $96\times96$ with periodic boundary condition. More
line scans are provided in the Supplementary Material \cite{Note2}.}
\end{figure*}

\subsection{PDW2}

We start with the uniform $s'\pm id$-wave PDW2 system {[}Eq.\,\eqref{eq:F_2},
$\gamma_{3}^{\Delta}<0${]}. In particular, we choose to add the $\Delta_{\pm Q_{y}}$
PDW orders and the results of selected momentum line scans {[}see
Fig.\,\ref{fig:BZ}{]} are shown in Fig.\,\ref{fig:PDW2-ARPES}(a).
More line scans are shown in Fig.\,5 of SM \cite{Note2}. We first
notice that the Fermi surface is gapped at both the antinodal regions
($k_{x,y}=\pi$ scans), despite that we have only added the PDW pairings
centered at $\pm Q_{y}/2$. The spectra near the nodal regions ($k_{x,y}=\frac{\pi}{2}$
scans) are gapless and resemble the bare spectra without the PDW pairing,
except the existence of some gap structures away from the Fermi energy.
These results constitute the Fermi arcs observed in the ARPES experiments
\cite{Norman1998,Comin24012014}. Within the PDW picture, these Fermi
surface segments are due to poor pairing condition near the Fermi
surface. However, the ``gap,'' defined as the energy gap at the
back-bending momentum $k_{G}$ below the chemical potential, is not
monotonically decreasing from the antinodes {[}compare Fig.\,5(a-IV)
and (a-V) in the SM \cite{Note2}{]}. This is however inconsistent
with the ARPES \cite{Lee2007a,Kondo2009}. Next, the band structures
near two antinodes are different. Near $\pm Q_{x}/2$ {[}Fig.\,\ref{fig:PDW2-ARPES}(a-III){]},
the spectrum resembles closely a conventional band structure gapped
by an order parameter such that the lower band edge at $\sim-0.2\thinspace\mathrm{eV}$
is left below the Fermi energy. Near $\pm Q_{y}/2$ {[}Fig.\,\ref{fig:PDW2-ARPES}(a-I){]},
where the PDW order is centered, we notice the so-called $k_{F}$-$k_{G}$
misalignment of the band gap, where $k_{F}$ is the Fermi momentum
of gapless bare bands (white dash lines) at high enough temperature.
This misalignment observed in the ARPES experiments imposes a strong
constraint on theories. Although this uniform PDW2 state can explain
several important features observed in ARPES, but more satisfactory
results could be obtained for the systems we will consider below.

Next we turn to the $s'\pm id$-wave PDW2 system with RFD induced
random phases. We set the amplitude $|\Delta_{\pm Q_{y}}(R)|$ to
be constant and use the averaged phases in Fig.\,\ref{fig:pdw2-cdw}
obtained by MC simulation as input to compute the spectral functions.
The results with the same line scans are shown in Fig.\,\ref{fig:PDW2-ARPES}(b).
The spectra are similar to those in the uniform PDW2 state, except
a noticeable difference that the coherent peaks are more broadened
near the pairing centers $\pm Q_{y}/2$ than at the other antinodal
region $\pm Q_{x}/2$ {[}compare Fig.\,\ref{fig:PDW2-ARPES}(b-I)
and (b-III){]}. Near the nodal regions, the spectra are largely unaltered
by the random phases, except some minor spreading of the spectral
weights. So the main effect of random phases is broadening of the
coherent peaks near the pairing centers and the band structure induced
by the PDW orders is largely preserved.

\begin{figure*}
\begin{centering}
\includegraphics[width=0.9\textwidth]{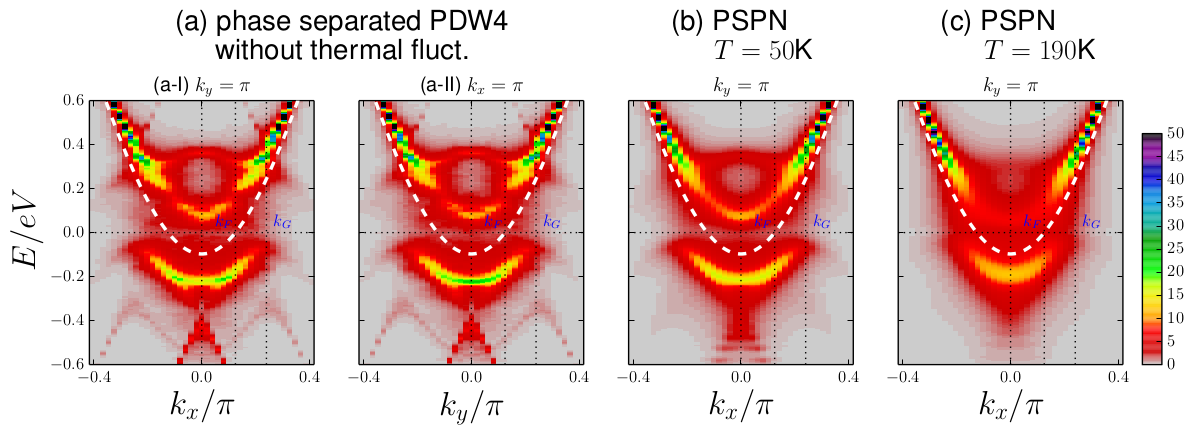}
\par\end{centering}

\caption{\label{fig:PDW4-ARPES}(Color online) Spectral functions for (a) phase-separated
PDW4 without thermal fluctuations, (b) PSPN state with thermal fluctuations
at $T=50\thinspace\mathrm{K}<T_{CO}$, and (c) PSPN state at $T=190\thinspace\mathrm{K}>T_{CO}$
are plotted at $k_{x,y}=\pi$. The PDW orders $\Delta_{\pm Q_{x,y}}$
are added at four pairing centers $\pm Q_{x,y}/2$. In panel (a),
we take the averaged PDW amplitude results $|\Delta_{\pm Q_{x,y}}|$
from Fig.\,\ref{fig:pdw4} as input and the PDW phases are homogeneous
and constant. The spectra closely resemble superposition of two PDW2
spectra with either $\Delta_{\pm Q_{x}}$ or $\Delta_{\pm Q_{y}}$
and the fourfold rotational symmetry is approximately preserved. In
panels (b) and (c), we have performed the thermal and disorder averages
at the stated temperatures. We note that the spectrum superposition
feature revealed in (a) remains. In all cases, the $k_{F}$-$k_{G}$
misalignment feature is intact, though the broadening is more severe
in the two high temperature cases. The parameters are the same as
in Fig.\,\ref{fig:PDW2-ARPES}. More line scans are provided in the
Supplementary Material \cite{Note2}.}
\end{figure*}

\subsection{phase-separated PDW4}

We now discuss about the $s^{\prime}\pm id$-wave PDW4 state with
the $\Delta_{x,y}$-phase separation {[}Eq.\,\eqref{eq:F4}, $\gamma_{2,3}^{\Delta}>0${]}.
Firstly, we notice that the PDW2 spectra in Fig.\,\ref{fig:PDW2-ARPES},
especially near the antinodes, do not preserve the fourfold rotational
symmetry, while this symmetry appears to be unspoiled in the ARPES
results. In the following, we show that the RFD induced phase separation
pattern for PDW4 can help to resolve this discrepancy, namely approximately
preserving the fourfold rotational symmetry while still maintaining
all the spectral features obtained in PDW2.

Here the PDW orders exists at all four momenta $\pm Q_{x,y}$. We
take the averaged $\Delta_{x,y}$ amplitudes with phase separation
in Fig.\,\ref{fig:pdw4} as input to calculate the spectral functions.
Also focusing on the effect of phase separation, we set the PDW phases
as spatial independent (i.e.\,constant). The results of line scans
$k_{x,y}=\pi$ are shown in Fig.\,\ref{fig:PDW4-ARPES}(a) and more
scans are presented in Fig.\,6 of the SM \cite{Note2}. PDW orders
now exist at all four pairing centers $\pm Q_{x,y}/2$, the phase
separated orders result in spectrum broadening near both antinodal
regions. The effect of amplitude fluctuations in PDW4 is similar to
that of phase fluctuations in PDW2 above in terms of spectrum broadening,
except it affects both antinodal regions. The nodal Fermi surface
is again gapless with minor peak broadening {[}Fig.\,6(a-VII) and
(a-XIV) in SM \cite{Note2}{]}. Next we observe that the spectra appear
as superposition of those from a PDW2 with $\Delta_{\pm Q_{x}}$ orders
and another one with $\Delta_{\pm Q_{y}}$ orders (see Fig.\,7 in
SM \cite{Note2} for detailed discussion) and thus approximately preserve
the fourfold rotational symmetry. This is an intriguing result since
the RFD induces directional PDW domains breaking the local fourfold
rotational symmetry. We note that the spectra at $k_{x,y}=\pi$ {[}Fig.\,\ref{fig:PDW4-ARPES}(a){]}
actually agree with the ARPES results \cite{Hashimoto2010,He25032011}
better than the PDW2's {[}Fig.\,\ref{fig:PDW2-ARPES}(a-I){]}, in
which the lower band edge near $\sim-0.2\thinspace\mathrm{eV}$ is
missing. We also note that the dispersion approaches Fermi level from
below when moving from antinodes towards the nodal regions {[}see
Fig.\,6(a) and Fig.\,8 in SM \cite{Note2}{]}, consistent with ARPES
results \cite{Yang2008,PhysRevLett.107.047003,He25032011}. Near the
end of the gapless Fermi arc (see Fig.\,8 in SM \cite{Note2}), the
dispersion is not particle-hole symmetric but bends downward and loses
spectral weight, qualitatively consistent with the ARPES results by
Yang \textit{et\,al.}\,\cite{Yang2008,PhysRevLett.107.047003}.
The same data, as well as some antinodal features, has also been nicely
captured by the phenomenological YRZ model \cite{Yang2006,PhysRevLett.107.047003,LeBlanc2011,Rice2012}.
We have also performed a similar calculation on the PDW4 with coexisting
$\Delta_{x,y}$ ($\gamma_{2}^{\Delta}<0$), the spectrum does not
have the superposition feature and fails to reproduce the ARPES results
due to the strong interference effect of the $Q_{x}$ and $Q_{y}$
PDW orders.

\subsection{PSPN state with (thermal) phase fluctuations}

We are finally ready to discuss the spectral function results for
the PSPN state with phase fluctuations. This state stems from the
$s^{\prime}\pm id$-wave PDW4 state. In order to capture the phase
fluctuations of the PDW orders, it is insufficient to compute only
the static averaged configurations as in above. Actually we need to
compute the averaged qausiparticle spectral function over thermal
fluctuations, given by \cite{Han2010} 
\begin{equation}
\left\langle A(k,\omega)\right\rangle =\frac{\sum_{\left\{ \Delta\right\} }A_{\left\{ \Delta\right\} }\left(k,\omega\right)e^{-\frac{1}{k_{B}T}F_{4,\mathrm{lattice}}}}{\sum_{\left\{ \Delta\right\} }e^{-\frac{1}{k_{B}T}F_{4,\mathrm{lattice}}}}\label{eq:MC-SpecFun}
\end{equation}
where for brevity $\left\{ \Delta\right\} $ denotes $\left\{ \Delta_{sQ_{a}}\right\} $.
This averaged spectral function $\left\langle A(k,\omega)\right\rangle $
is calculated by taking the thermal average with $\left\{ \Delta_{sQ_{a}}\right\} $
configurations generated by MC simulations of the GL functional $F_{4,\mathrm{lattice}}$
in Eq.\,\eqref{eq:F4}. We then perform the disorder average over
$16$ RFD configurations. We carry out the numerical calculations
at two temperatures $T=50\thinspace\mathrm{K}$ and $T=190\thinspace\mathrm{K}$,
which are at opposite sides of the charge order crossover temperature
of $T_{CO}\sim150\thinspace\mathrm{K}$. Again, for each RFD configuration,
we take the amplitudes from the MC results with both amplitude and
phase updates and simulated annealing process as in Fig.\,\ref{fig:pdw4}
as input and then carry out the MC simulations on the PDW phases at
the desired temperatures. The results for $50\thinspace\mathrm{K}$
and $190\thinspace\mathrm{K}$ are respectively presented in Fig.\,\ref{fig:PDW4-ARPES}(b)
and (c). The spectra are smoothened compared with those without phase
fluctuations, but the band structures are largely unaltered compared
with the PDW4 results without phase fluctuations {[}Fig.\,\ref{fig:PDW4-ARPES}(a){]}.
Many important features that we obtained previously remain intact
even with the presence of the phase fluctuations. Firstly, the ``Fermi
arcs'' remain. The spectra near the antinodes are severely broadened.
Compared with the PDW4 results without phase fluctuations {[}Fig.\,\ref{fig:PDW4-ARPES}(a){]},
the broadening mainly comes from the $\Delta_{x,y}$-phase separation
instead of the phase fluctuations. The ``gap'' decreases monotonically
away from the antinodes owing to the phase fluctuation induced gap
fillings and finally becomes gapless near the nodal regions. Again
away from antinodes, the spectrum approaches the Fermi level from
below, in accord with ARPES experiments \cite{Yang2008,PhysRevLett.107.047003,He25032011}.
Secondly, the band structure superposition and the $k_{F}$-$k_{G}$
misalignment also survive the phase fluctuations. Moreover, if we
compare the results for $50\thinspace\mathrm{K}$ and $190\thinspace\mathrm{K}$,
we see that the crossover temperature $T_{CO}$ does not induce a
qualitative change on the spectrum. The two major features mentioned
above still survive the depinning of the smectic phases across $T_{CO}$
and can be regarded as definitive characteristics of the PDW orders
for $T<T^{*}$. These results agree well with the ARPES measurements
\cite{Lee2007a,Kondo2009,Hashimoto2010,He25032011,Comin24012014,Hashimoto2014,Vishik2010},
in which also show these features in the pseudogap region below $T^{*}$.
(We note that these ARPES features can also be captured to some extent
by alternative theories \cite{PhysRevB.83.212503,Wang2015a}) Comparison
of the result and the ARPES data is given in Fig.\,9 of the SM \cite{Note2}.

By comparing the spectral results for different PDW states, we notice
two major effects of RFD. Firstly, in terms of the phase separation
pattern and the random phases, it broadens the spectrum, as one would
have expected from a disordered system. The broadening is more severe
near the antinodes, which is also observed in ARPES data \cite{Orenstein21042000}.
But more importantly, RFD induces the phase separation pattern for
the directional orders, rendering the spectra approximately obeying
the fourfold symmetry, while still preserving many experimentally
observed features appeared in the directional PDW2 state.

Lastly we argue that the superposition feature actually support the
proposal that the cuprate SC has stripe PDW and CDW orders ($\gamma_{2}^{\Delta}>0$).
Since without PDW order, the fermionic spectrum in the CDW phase cannot
reproduce the ARPES results \cite{Yang2008,PhysRevLett.107.047003,He25032011,PhysRevX.4.031017}
that the spectrum approaches from Fermi level from below, we shall
confine ourselves to various scenarios of PDW theories. First, the
pure PDW2 could not reproduce the lower band edge of the ARPES spectrum
as we discussed above. Next, we have also checked that the checkerboard
PDW4 state with $\gamma_{2,3}^{\Delta}<0$ could not reproduce the
ARPES results. A third possibility is proposed in Ref.\,\cite{PhysRevX.4.031017}
that the ARPES results actually constitute from a checkerboard PDW4
orders decaying in the momentum space. It is proposed that the PDW
symmetry factor $f_{sQ_{a}}(k)\sim e^{-k^{2}/\xi^{2}}$ decays away
from the pairing centers, and it can successfully explain the $k_{F}$-$k_{G}$
misalignment and the Fermi arcs. But the spectrum is largely coming
from the PDW2 states at the pairing centers due to the exponential
decaying SC symmetry factor, so it has missed the ARPES feature at
the lower band edge shown in Fig.\,\ref{fig:PDW4-ARPES}(a-I). Nevertheless,
the stripe PDW does not necessarily need to exist within the same
layer. Actually, alternating layers of $x$- and $y$-directional
stripe PDW \cite{Berg2007} would produce a similar superposition
spectra. But the phase separation structure similar to those in Fig.\,\ref{fig:pdw4}
is indeed observed in STM \cite{Mesaros22072011,Hamidian2015} and
also supported by REXS \cite{Comin20032015}, so it is perhaps favored
over the alternating layer scenario. It is hard to exhaust all the
possibilities, but it is intriguing to see the convergence of the
STM, REXS (indicating stripe CDW, induced by the stripe PDW) and the
ARPES (indicating stripe PDW) experimental results towards a unifying
theoretical framework.

\section{Conclusions}

In conclusion, the phase separated pair nematic (PSPN) state derived
from a four-component pair density wave (PDW) model under the influence
of random field disorders (RFD) captures coherently a number of experimental
results in the pseudogap. We consider, to the best of our knowledge,
a \textit{new} parameter regime $\gamma_{2,3}^{\Delta}>0$ for the
phenomenology of cuprates, in which all the PDW components $\Delta_{\pm Q_{a}}$
are competing with each other. The new choice of parameter impedes
the generalization of existing PDW results to the present case, however,
we show that in terms of a phenomenological GL approach, the experimental
observed features obtained in the existing PDW theory are retained
if disorders are included. It is shown that CDW and a time-reversal
symmetry breaking (TRSB) order in the form of loop current (LC) are
induced as secondary composite orders. The inclusion of disorders
permits the state to be spatially inhomogeneous (in terms of the PDW
components) and this readily resolves an issue concerning the incompatibility
of CDW and loop current order within a four-component PDW model \cite{Agterberg2015}.
Nonetheless, the CDW is short ranged under the influence of random
field disorders, consistent with the STM \cite{Mesaros22072011,Fujita2014,Hamidian2015}
and REXS \cite{Comin20032015} results, while the LC order can account
for the TRSB order observed in polar Kerr rotation \cite{Xia2008,Karapetyan2014}
and polarized neutron diffraction \cite{Fauque2006,Li2008,PhysRevLett.105.027004,Mangin-Thro2015}
experiments. Furthermore, by a MC simulation, we show that random
field disorders can also induce a phase separation pattern for the
CDW, similar to that observed in STM. This PSPN state also appears
to capture other experimental features. It explains a distinct temperature
scale $T_{CO}<T^{*}$ for the static (short range) CDW (see section
\ref{sec:PDW4}). More importantly, it accounts for the same pseudogap
temperature $T^{*}$ observed in linear resistivity, ARPES (for the
anomalous spectral features), REXS (for the dynamically fluctuating
directional CDW) and polarized neutron diffraction (for the IUC TRSB
order) experiments. We also argue that the thermal superconducting
phase fluctuations lead to finite flux flow resistance, resulting
in a non-superconducting state regardless of $T_{CO}$. To test the
proposed PDW theory, one may probe the half SC flux trapped at the
$2\pi$-CDW dislocation (see section \ref{sub:Topo-defects} and \ref{sub:PSPN-numerics}),
provided that the local SC phase is strongly pinned. In section \ref{sec:Constraint},
we constrain the PDW, from the observed CDW with a predominantly $d$-wave
form factor, to have $s^{\prime}\pm id$ pairing symmetries. In section
\ref{sec:ARPES}, we show that a number of anomalous features in ARPES,
namely Fermi arcs, $k_{F}$-$k_{G}$ misalignment, antinodal gap closing
from below are retained in the PSPN state, even with thermal fluctuations.
Moreover, the PSPN state is shown to have severe coherent peak broadening
near the antinodes (but not the nodes), which has been a puzzling
feature observed in ARPES \cite{Orenstein21042000}. The random field
disorder induced phase separation pattern helps to explain why the
ARPES spectrum still obey the fourfold rotational symmetry approximately,
while locally it has been broken as revealed in STM. From above, we
see that the random field disorders assist in understanding the pseudogap
for $T<T^{*}$ in terms of the PSPN state by resolving several issues
in previous studies.
\begin{acknowledgments}
C.C.\,acknowledges useful discussions with Zheng-Xin Liu, Luyang
Wang and Zi-Xiang Li and helpful comments from Marc-Henri Julien.
The author would also like to thank Hong Yao for previous collaboration
on a related topic and Patrick A.\ Lee for his stimulating talk at
the HKUST IAS 2014 program ``topological matter, superconductivity
and Majorana,'' which inspired the present work. Last but not least,
the criticisms, suggestions and comments raised by the anonymous referees,
especially the ``First Referee,'' for improving the manuscript to
the current form are highly appreciated.
\end{acknowledgments}

\begin{table}
\begin{tabular}{|c|>{\raggedright}m{0.65\columnwidth}|}
\hline 
Acronym & Full phrase\tabularnewline
\hline 
\hline 
ARPES & Angular resolved photoemission spectroscopy\tabularnewline
\hline 
BZ & Brillouin zone\tabularnewline
\hline 
CDW & Charge density wave\tabularnewline
\hline 
dSC & $d$-wave SC\tabularnewline
\hline 
IUC & Intra-unit cell\tabularnewline
\hline 
LC & Loop current\tabularnewline
\hline 
MC & Monte Carlo\tabularnewline
\hline 
NMR & Nuclear magnetic resonance\tabularnewline
\hline 
PDW & Pair density wave\tabularnewline
\hline 
PDW2 & Two-component PDW\tabularnewline
\hline 
PDW4 & Four-component PDW\tabularnewline
\hline 
PKR & Polar Kerr rotation\tabularnewline
\hline 
PND & Polarized neutron diffraction\tabularnewline
\hline 
PSPN & Phase separated pair nematic\tabularnewline
\hline 
REXS & Resonant inelastic X-ray scattering\tabularnewline
\hline 
RIXS & Resonant elastic X-ray scattering\tabularnewline
\hline 
RFD & Random field disorder\tabularnewline
\hline 
SC & Superconductor / superconductivity / superconducting\tabularnewline
\hline 
SM & Supplementary material (Ref.\,\cite{Note2})\tabularnewline
\hline 
STM & Scanning tunneling microscopy\tabularnewline
\hline 
TRSB & Time reversal symmetry breaking\tabularnewline
\hline 
\end{tabular}

\caption{Acronym list for the phrases commonly used in the paper.}
\end{table}

\appendix

\section{CDW form factor induced by $s'\pm id$ PDW on a square lattice\label{sec:CDW-form-factor}}

From the definitions of PDW Eq.\,\eqref{eq:Delta_def} with symmetry
factor Eq.\,\eqref{eq:pair_sym_cont}
\begin{eqnarray*}
f_{+Q}(k) & = & (k_{x}^{2}+k_{y}^{2})+\alpha(k_{x}^{2}-k_{y}^{2})\\
f_{-Q}(k) & = & f_{+Q}^{*}(k)
\end{eqnarray*}
(here $\alpha\in\mathbb{C}$) and its induced CDW
\[
\rho\left(r,+2Q\right)=\Delta^{*}\left(r,-Q\right)\Delta\left(r,+Q\right)\:,
\]
we have 
\begin{eqnarray*}
\rho(r,2Q) & \propto & \sum_{k_{1}k_{2}}f_{+Q}(k_{1})f_{+Q}(k_{2})e^{-ik_{1}r}e^{+ik_{2}r}\\
 & = & \sum_{p}\rho_{2Q}(p)e^{-ipr}\thinspace,
\end{eqnarray*}
and 
\[
\rho_{2Q}(p)=\int_{0}^{2\pi}d\theta_{q}f_{+Q}(q+p/2)f_{+Q}(q-p/2)
\]
where $\theta_{q}=\arg q$, and we define $k_{1}=q+\frac{p}{2}$ and
$k_{2}=q-\frac{p}{2}$. After some straightforward calculations, we
have
\begin{eqnarray*}
\rho_{2Q}(p) & = & \frac{1}{8\pi}[(1+\alpha^{2}/2)(p_{x}^{2}+p_{y}^{2})^{2}\\
 &  & \qquad+2\alpha(p_{x}^{2}+p_{y}^{2})(p_{x}^{2}-p_{y}^{2})\\
 &  & \qquad+(\alpha^{2}/2)(p_{x}^{4}-6p_{x}^{2}p_{y}^{2}+p_{y}^{4})]+\mathcal{O}(p^{2})
\end{eqnarray*}
We note that the first term is clearly $s$-wave. While the second
(third) term is $d$-wave ($g$-wave) due to the identities $\cos2\theta_{p}=\frac{p_{x}^{2}-p_{y}^{2}}{p_{x}^{2}+p_{y}^{2}}$
and $\cos4\theta_{p}=\frac{p_{x}^{4}-6p_{x}^{2}p_{y}^{2}+p_{y}^{4}}{(p_{x}^{2}+p_{y}^{2})^{2}}$.
A less rigorous derivation by the product of the basis functions
\begin{eqnarray*}
\rho_{2Q}(p) & \sim & (1+\alpha\cos2\theta_{p})^{2}\\
 & = & (1+\alpha^{2}/2)+2\alpha\cos2\theta_{p}+(\alpha^{2}/2)\cos4\theta_{p}
\end{eqnarray*}
would also give the same result.

Similarly, in the lattice case, we have symmetry factor Eq.\,\eqref{eq:pair_sym}
\[
f(\vec{k})=\left(\cos k_{x}+\cos k_{y}\right)+i\sqrt{2}\left(\cos k_{x}-\cos k_{y}\right)\thinspace.
\]
Then 
\begin{eqnarray*}
\rho(p,2Q) & = & \sum_{q}\left[\left(-1+2\sqrt{2}i\right)\cos k_{1x}\cos k_{2x}\right.\\
 &  & +\left(-1-2\sqrt{2}i\right)\cos k_{1y}\cos k_{2y}\\
 &  & +3\cos k_{1x}\cos k_{2y}+3\cos k_{1y}\cos k_{2x}\Bigr]\thinspace.
\end{eqnarray*}
Then we substitute $k_{1}=q+\frac{p}{2}$, $k_{2}=q-\frac{p}{2}$
and evaluate the integrals $\sum_{q}\rightarrow\int_{-\pi}^{\pi}dq_{x}\int_{-\pi}^{\pi}dq_{y}$,
we have 
\begin{eqnarray*}
\rho(p,2Q) & = & +\Delta_{d}^{\mathrm{lat}}\left(\cos p_{x}-\cos p_{y}\right)\\
 &  & +\Delta_{g}^{\mathrm{lat}}\left(\cos p_{x}+\cos p_{y}\right)\thinspace,
\end{eqnarray*}
where $\Delta_{d}^{\mathrm{lat}}=4\sqrt{2}\pi^{2}i$ and $\Delta_{g}^{\mathrm{lat}}=-2\pi^{2}$.
In the continuous case, after setting the $s'$-wave component to
zero, we have
\[
\rho_{2Q}(k)=2\sqrt{2}i\Delta_{s}^{2}(k_{x}^{4}-k_{y}^{2})-\Delta_{s}^{2}(k_{x}^{4}-6k_{x}^{2}k_{y}^{2}+k_{y}^{4})\thinspace.
\]
The ratio of the $d$-wave to $g$-wave component is $-2\sqrt{2}i$,
matches that of the lattice case $\Delta_{d}^{\mathrm{lat}}/\Delta_{g}^{\mathrm{lat}}=-2\sqrt{2}i$.
We note that although the \textit{induced} CDW's form factor has a
``$g$-wave'' component, it manifests as $s'$-wave $\sim\cos p_{x}+\cos p_{y}$
due to the choice of nearest neighbor bond order and the square lattice.

\bibliographystyle{apsrev4-1}
\bibliography{refs}

\end{document}


\title{Supplemental Material\\
Interplay between pair density waves and random field disorders in
the pseudogap regime of cuprate superconductors}

\author{Cheung Chan}

\affiliation{Institute for Advanced Study, Tsinghua University, Beijing, 100084,
China}

\maketitle
\vspace{1cm}

\section{Short range directional CDW and charge-4e superconducting correlations
in PSPN state\label{sec:discuss}}

In this section, we discuss the correlations of directional CDW and
charge-4e superconducting order in the phase-separated pair nematic
(PSPN) state. In section \ref{sub:PDW2}, we shall briefly review
these correlations and the corresponding phases in the two-component
PDW system. Next in section \ref{sub:PDW4}, we generalize the correlations
to the four-component PDW system and provide numerical evidence for
the short range correlations as argued in the main text. In section
\ref{sub:gamma4}, we discuss the rationale behind the parameter choice
$\gamma_{4}^{\Delta}=0$ and its implications to the PSPN state.

\subsection{Brief review on two-component PDW model\label{sub:PDW2}}

Here we briefly summarize the possible states derived from a two-component
PDW (PDW2) system with components $\Delta_{\pm Q}(r)$. We define
the composite orders, namely the CDW 
\begin{equation}
\rho_{2Q}(r)=\Delta_{-Q}^{*}(r)\Delta_{+Q}(r)
\end{equation}
 and the charge-4e superconductivity
\begin{equation}
\Delta_{4e}(r)=\Delta_{-Q}(r)\Delta_{+Q}(r)
\end{equation}
As briefly discussed in the section II.1 in the main text, there are
four possible states induced by proliferating different topological
defects provided that the local PDW order $|\Delta(r)|=\sqrt{\sum_{s}|\Delta_{sQ}(r)|^{2}}$
is non-vanishing (see for example Ref.\,\cite{Berg2009b,Radzihovsky2011}
for more details). The proliferation of SC $2\pi$-vortices destroys
the superconductivity, resulting in a (pure) CDW state. In this CDW
state, the directional CDW correlations 
\begin{equation}
\tilde{C}_{\rho,2Q}(r-r')=\langle\rho_{2Q}^{*}(r)\rho_{2Q}(r')\rangle
\end{equation}
 are long ranged and the charge-4e SC correlation 
\begin{equation}
\tilde{C}_{\Delta,4e}(r-r')=\langle\Delta_{4e}^{*}(r)\Delta_{4e}(r')\rangle
\end{equation}
is short ranged (exponential decay). We note that the short range
charge-4e SC correlation is induced by the proliferation of SC $2\pi$-vortices,
as in the Berezinsky- Kosterlitz-Thouless (BKT) transition. We summarize
the states, the corresponding topological defects proliferated, and
the correlations in Table \ref{tab:FourStates} as the ease of reference.
In particular, we shall focus on the nematic state with short range
$\tilde{C}_{\rho,2Q}(r)$ and $\tilde{C}_{\Delta,4e}(r)$ is induced
by proliferation of the half-vortices, which would manifest as CDW
$2\pi$-dislocations in $\rho_{2Q}$ and charge-4e SC $2\pi$-vortices
in $\Delta_{4e}$ (see Table \ref{tab:TopoDefects} for more details).

\subsection{Four-component PDW model and PSPN state\label{sub:PDW4}}

In the four-component PDW system, we generalize the PDW2 case to define
the composite orders, namely the directional CDW 
\begin{equation}
\rho_{2Q_{a}}(r)=\Delta_{-Q_{a}}^{*}(r)\Delta_{+Q_{a}}(r)
\end{equation}
 ($a=x,y$) and charge-4e superconductivity 
\begin{equation}
\Delta_{4e}(r)=\sum_{a=x,y}\Delta_{4e}^{a}(r)=\sum_{a}\Delta_{-Q_{a}}(r)\Delta_{+Q_{a}}(r)\:.\label{eq:totalchrg4e}
\end{equation}
We note that since only phase separated directional PDW components
$\Delta_{a}=\sqrt{|\Delta_{+Q_{a}}|^{2}+|\Delta_{-Q_{a}}|^{2}}$ are
considered here, so we simply add the directional charge-4e order
parameters 
\begin{equation}
\Delta_{4e}^{a}(r)=\Delta_{-Q_{a}}(r)\Delta_{+Q_{a}}(r)\:,
\end{equation}
to give $\Delta_{4e}$ as above. Similar to the PDW2 case, there are
four possible states induced by proliferating different topological
defects provided that the local PDW order $|\Delta(r)|=\sqrt{\sum_{sa}|\Delta_{sQ_{a}}(r)|^{2}}$
is non-vanishing. 

For the PDW4 system, we similarly define the directional CDW correlations
\begin{equation}
C_{\rho,2Q_{a}}(r-r')=\langle\!\langle\rho_{2Q_{a}}^{*}(r)\rho_{2Q_{a}}(r')\rangle\!\rangle
\end{equation}
and the charge-4e SC correlation
\begin{equation}
C_{\Delta,4e}(r-r')=\langle\!\langle\Delta_{4e}^{*}(r)\Delta_{4e}(r')\rangle\!\rangle\thinspace.
\end{equation}
Here $\langle\!\langle\cdot\rangle\!\rangle$ denotes thermal and
disorder averages. Depending on the behaviors of these correlations
we can also have four possible states as in the PDW2 case. In particular,
we shall focus on the nematic state with short range $C_{\rho,2Q_{a}}(r)$
and $C_{\Delta,4e}(r)$ is induced by proliferation of the half-vortices,
which would manifest as CDW $2\pi$-dislocations in $\rho_{2Q_{a}}$
and charge-4e SC $2\pi$-vortices in $\Delta_{4e}^{a}$.

\begin{figure}[H]
\begin{centering}
\includegraphics[width=0.7\textwidth]{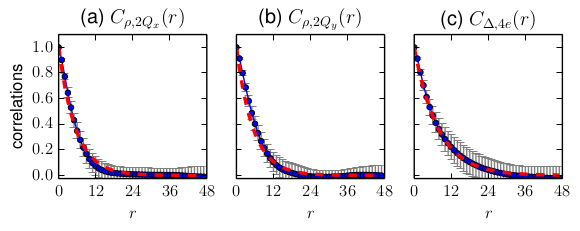}
\par\end{centering}

\caption{\label{fig:Correlations}Spatial correlations of (a,b) directional
CDW $\rho_{2Q_{a}}=\Delta_{-Q_{a}}^{*}\Delta_{+Q_{a}}$ ($a=x,y$)
and (c) charge-4e superconductivity $\Delta_{4e}=\sum_{a=x,y}\Delta_{-Q_{a}}\Delta_{+Q_{a}}$
of the PSPN state at $T=10$\,K. The results are averaged over 80
RFD configurations (disorder average). The blue dots are the mean
values of the correlations and the error bars are for one standard
deviation. The red dashed lines are the exponential fits for each
set of data. The results show that the PSPN state has both short range
CDW and charge-4e SC correlations and it is nematic (locally). See
the discussions in section \ref{sec:discuss} above for the meaning
of the correlations, corresponding states and the MC process.}
\end{figure}

Next, we describe the MC process to compute the correlations under
the influence of the PDW phase fluctuations. For each RFD configuration,
we first perform a simulated annealing down to $T=10$\,K with both
amplitude and phase updates (as in stage one described in section
III.1 in the main text) to obtain the averaged PDW orders. We then
use the averaged PDW orders to perform a MC simulations and measure
various correlations shown in Fig.\,\ref{fig:Correlations} with
only PDW phase updates (as in stage two described in section III.2
in the main text). Finally, we average the results over a number of
RFD configurations to obtain the disorder average.

The results on the $C_{\rho,2Q_{a}}(r)$ and $C_{\Delta,4e}(r)$ are
shown in Fig.\,\ref{fig:Correlations}. It is clear that both correlations
decay exponentially and are short ranged, resulting in a nematic state
(at least locally within each directional PDW domain). These correlations
are actually induced by the half-vortices in the directional PDW,
which manifest as CDW $2\pi$-dislocations in $\rho_{2Q_{a}}$ {[}see
Fig.\,5(a-III) and (b-III) in the main text{]} and charge-4e SC $2\pi$-vortices
in $\Delta_{4e}^{a}$ (see Fig.\,\ref{fig:avg_chrg4e}).

\subsection{Rationale for choosing $\gamma_{4}^{\Delta}=0$\label{sub:gamma4}}

\begin{figure}[H]
\begin{centering}
\includegraphics[width=0.45\textwidth]{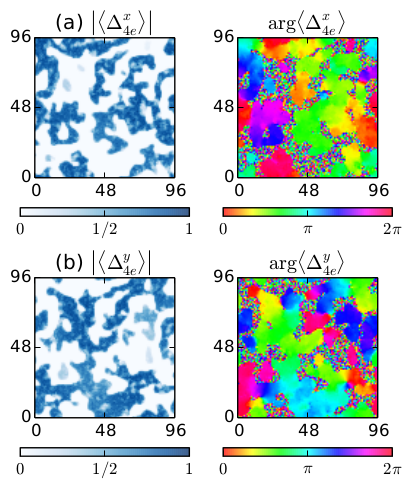}
\par\end{centering}

\caption{\label{fig:avg_chrg4e}Spatial variations of averaged directional
charge-4e superconducting orders $\langle\Delta_{4e}^{a}\rangle$
measured at $T=10$\,K, where $\Delta_{4e}^{a}(r)=\Delta_{-Q_{a}}(r)\Delta_{+Q_{a}}(r)$.
It is obvious that phase separation also occurs in the directional
components $\Delta_{4e}^{x,y}$, inherited from the directional PDW
orders. It is also evident that the phases of directional $\Delta_{4e}^{a}$
are not coupled due to the choice of $\gamma_{4}^{\Delta}=0$. From
the averaged orders $\Delta_{4e}^{x,y}$ at low temperature, we see
that static charge-4e SC $2\pi$-vortices is present owing to the
pinning of the PDW indirectly through the coupling between RFD and
CDW. This result can actually be inferred from the averaged PDW orders
$\langle\Delta_{sQ_{a}}\rangle$ shown in Fig.\,5 of the main text.
These \textit{static} vortices lead to the short range charge-4e SC
correlation $C_{\Delta,4e}(r)$ shown in Fig.\,\ref{fig:Correlations}.
The MC process is the same as those in Fig.\,\ref{fig:Correlations},
except that only one RFD configuration is computed here.}
\end{figure}

We finally discuss the implication of the fact that the pseudogap
is non-superconducting on the parameter $\gamma_{4}^{\Delta}$ in
the GL functional Eq.\,(10) of the main text. For the ease of reference,
we repeat the $\gamma_{4}^{\Delta}$ term here 
\begin{equation}
+2\gamma_{4}^{\Delta}\sum_{r}\mathrm{Re}\left[\Delta_{+Q_{x}}(r)\Delta_{-Q_{x}}(r)\Delta_{+Q_{y}}^{*}(r)\Delta_{-Q_{y}}^{*}(r)\right]\:.
\end{equation}
We note that if we identify the directional charge-4e SC components
$\Delta_{4e}^{a}(r)=\Delta_{-Q_{a}}(r)\Delta_{+Q_{a}}(r)$, then the
$\gamma_{4}^{\Delta}$ term can be regarded as the Josephson junction
coupling between the local $\Delta_{4e}^{x}$ and $\Delta_{4e}^{y}$
orders. We shall consider the case $\gamma_{4}^{\Delta}$ is finite
and negative. Then we expect that the directional charge-4e SC orders
$\Delta_{4e}^{a}$ would attain phase coherence at low enough temperature,
even if their amplitudes $|\Delta_{4e}^{a}|$ are phase separated
as shown in Fig.\,\ref{fig:avg_chrg4e}. Notice that the total charge-4e
SC order Eq.\,\eqref{eq:totalchrg4e} is finite everywhere, since
we require the finiteness in the PDW orders. As in the case of BKT
transition, the phase coherence in $\Delta_{4e}$ then implies that
the system is superconducting below $T\apprle T_{4e}\sim\gamma_{4}^{\Delta}$
(i.e.\ a charge-4e superconductor, see e.g.\,Ref.\,\cite{Berg2009b}).
Experimentally, the pseudogap region in the temperature range $T_{c}<T<T^{*}$
is not superconducting, where $T_{c}$ is the transition temperature
for the $d$-wave SC. This fact suggests that $\gamma_{4}^{\Delta}$
is small, such that the charge-4e superconductor induced by the phase
separated PDW orders discussed above is hidden inside the $d$-wave
SC dome, i.e.\,$T_{4e}<T_{c}$. In the main text, we simply set $\gamma_{4}^{\Delta}$
to zero In this simplified case, the phase coherence between directional
charge-4e SC $\Delta_{4e}^{a}$ cannot build up owing to phase separation
pattern (see Fig.\,\ref{fig:avg_chrg4e}) and we expect the PSPN
state to be non-superconducting. The above discussion provides a rationale
for such a parameter choice.

\vspace{1cm}

\bibliographystyle{apsrev4-1}
\bibliography{refs-sm}

\begin{table}
\begin{centering}
\begin{tabular}{|c|c|c|c|}
\hline 
 & Type of topological defect proliferated & Charge-4e SC correlation & CDW correlation\tabularnewline
\hline 
\hline 
PDW & none & Long range & Long range\tabularnewline
\hline 
CDW & $\left(1,0\right)$: SC $2\pi$-vortex & Short range & Long range\tabularnewline
\hline 
Charge-4e SC & $\left(0,1\right)$: $2\pi$-dislocation & Long range & Short range\tabularnewline
\hline 
Nematic & $\left(\frac{1}{2},\frac{1}{2}\right)$: half vortex & Short range & Short range\tabularnewline
\hline 
\end{tabular}
\par\end{centering}

\caption{\label{tab:FourStates}List of four states accessible in a two-component
PDW (PDW2) system by proliferation of different types of topological
defects labeled by $(n_{v},n_{d})$ in Eq.\,(7) in the main text.}
\end{table}

\begin{table}
\begin{centering}
\begin{tabular}{|c|c|c|c|}
\hline 
PDW & $\left(\frac{1}{2},\frac{1}{2}\right)$: half vortex & $\left(1,0\right)$: SC $2\pi$-vortex & $\left(0,1\right)$: $2\pi$-dislocation\tabularnewline
\hline 
\hline 
CDW & CDW $2\pi$-dislocation  & -- & CDW $4\pi$-dislocation\tabularnewline
\hline 
Charge-4e SC & Charge-4e SC $2\pi$-vortex & Charge-4e SC $4\pi$-vortex & --\tabularnewline
\hline 
\end{tabular}
\par\end{centering}

\caption{\label{tab:TopoDefects}List of topological defects in a two-component
PDW (PDW2) system labeled by $(n_{v},n_{d})$ in Eq.\,(7) in the
main text. In the second and the third rows, we list the topological
defects in terms of the induced CDW and charge-4e superconductivity
manifested by the corresponding defects in the PDW.}
\end{table}

\vspace{5cm}

\appendix

\begin{figure}[H]
\begin{centering}
\includegraphics[width=0.75\columnwidth]{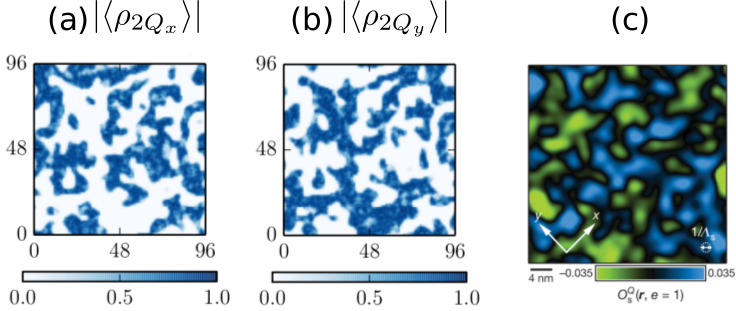}
\par\end{centering}

\caption{Comparison of the Monte Carlo (MC) result for the CDW (a) $|\langle\rho_{2Q_{x}}\rangle|$
and (b) $|\langle\rho_{2Q_{y}}\rangle|$ {[}taken from Fig.\,5(a-III)
and (b-III) in the main text{]} and (c) the \textit{smectic modulation
measure} $O_{s}^{Q}(e)=\bar{Z}\left(e\right)^{-1}\left[\mathrm{Re}Z(S_{y},e)-\mathrm{Re}Z(S_{x},e)\right]$
in the STM experiments \cite{Lawler2010,Mesaros22072011}, where we
shall identify $S_{x(y)}=2Q_{x(y)}$. Here $O_{s}^{Q}(e)$ signified
the relative strength of the local smectic (directional) CDW order
in the $x$- and $y$-directions. We note that the MC simulation largely
reproduces the STM results that show phase separation of $2Q_{x}$-
and $2Q_{y}$-CDW's. \textit{Note}: In Ref.\,\cite{Lawler2010} and
\cite{Mesaros22072011}, the authors defined another quantity, the
\textit{intra-unit-cell electronic nematicity}: $O_{n}^{Q}(e)=\bar{Z}(e)^{-1}\left[\mathrm{Re}Z(Q_{y}^{\prime},e)-\mathrm{Re}Z(Q_{x}^{\prime},e)\right]$,
where $Q_{x}^{\prime}=(2\pi/a_{0},0)$ and $Q_{y}^{\prime}=(0,2\pi/a_{0})$.
This quantity does not measure any smectic modulation since $Q_{x,y}^{\prime}$
is just the reciprocal lattice vectors (not to be confused with our
$Q_{x,y}$ for the PDW wave vectors).}
\end{figure}

\begin{figure}[H]
\begin{centering}
\includegraphics[width=0.35\paperwidth]{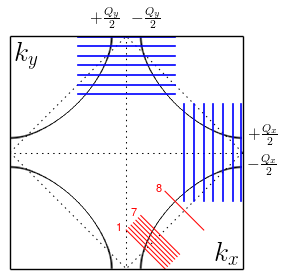}
\par\end{centering}

\caption{\label{fig:lineScans}Line scans in blue for the quasiparticle spectral
functions in Fig.\,\ref{fig:PDW2-SpecFun} and \ref{fig:PDW4-SpecFun}.
Those in red for Fig.\,\ref{fig:cmp-JS}.}
\end{figure}

\begin{figure}[H]
\centering{}\includegraphics[width=0.65\columnwidth]{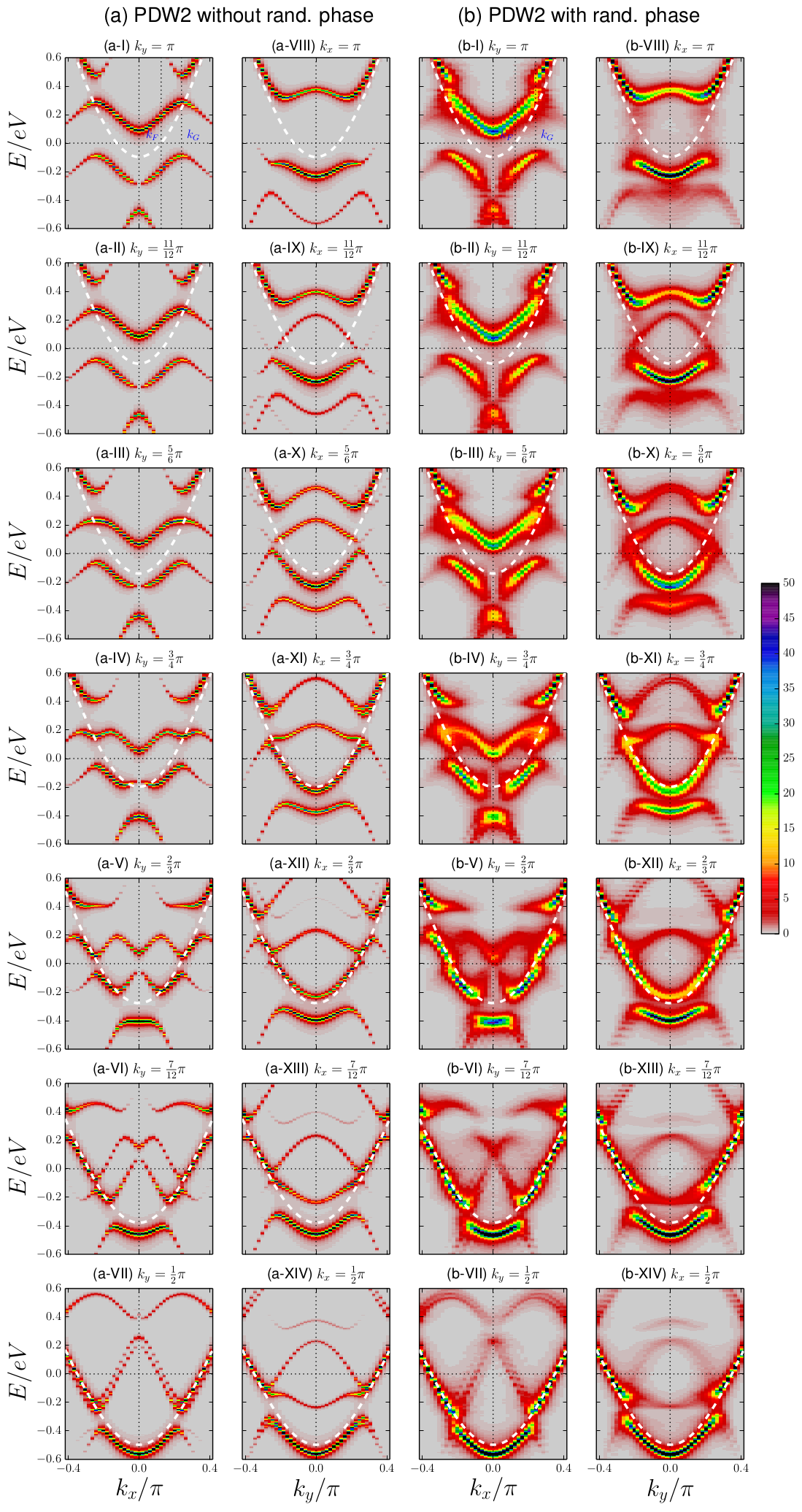}\caption{\label{fig:PDW2-SpecFun}The figure is an extended version for Fig.\,9
in the main text and more line scans (see Fig.\,\ref{fig:lineScans})
are presented. Panel (a) shows the PDW2 system without random phases
and panel (b) shows the PDW2 with static random phases taken from
Fig.\,3 in the main text.}
\end{figure}

\begin{figure}[H]
\centering{}\includegraphics[width=0.95\columnwidth]{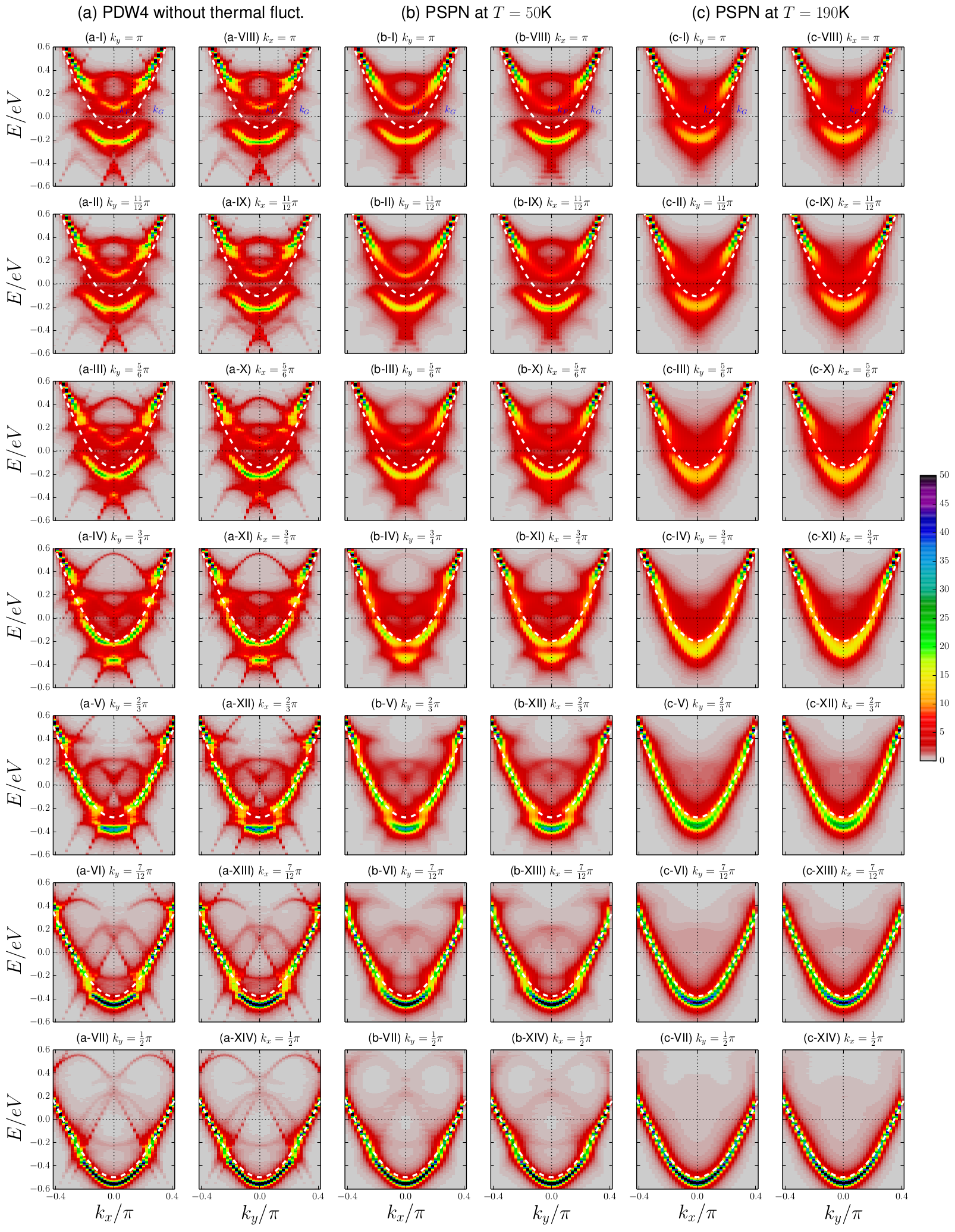}\caption{\label{fig:PDW4-SpecFun}The figure is an extended version for Fig.\,10
in the main text and more line scans (see Fig.\,\ref{fig:lineScans})
are presented. Spectral functions for (a) phase-separated PDW4 without
thermal fluctuations, (b) PSPN state with thermal fluctuations at
$T=50\thinspace\mathrm{K}<T_{CO}$, and (c) PSPN state at $T=190\thinspace\mathrm{K}>T_{CO}$.
We note the so-called ``Fermi arcs,'' that the spectra near the
anti-nodes are gapped. While moving towards the nodal regions the
gap diminishes monotonically and the spectrum eventually becomes gapless
and coincides with the bare band.}
\end{figure}

\begin{figure}[H]
\begin{centering}
\includegraphics[width=0.75\columnwidth]{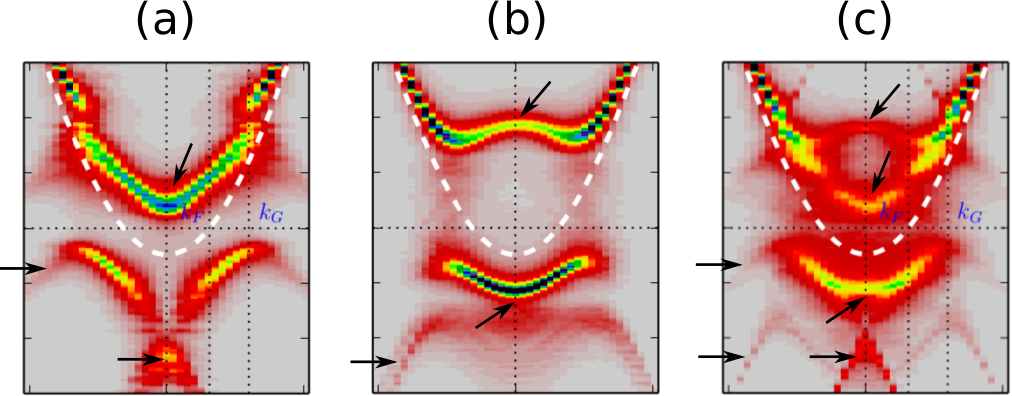}
\par\end{centering}

\caption{Spectral function superposition feature for the phase-separated PDW4
state. Panels (a) and (b) are the sepctral functions at the BZ boundaries
along $k_{y}=\pi$ and $k_{x}=\pi$ respectively. Panel (a) {[}(b){]}
is taken from Fig.\,\ref{fig:PDW2-SpecFun}(b-I) {[}Fig.\,\ref{fig:PDW2-SpecFun}(b-VIII){]}.
Panels (c) is for phase-separated PDW4 state, taken from Fig.\,\ref{fig:PDW4-SpecFun}(a-I).
We notice panel (c) resembles a spectral function superposed from
(a) and (b), by noticing that the various features presented in them
also appear in (c). We have marked these features in black arrows
for the ease of reference. We thus conclude that though the PDW4 state
is phase separated with $Q_{x}$ and $Q_{y}$ domains, its spectral
functions resemble superposition spectrum from those of $Q_{x}$-
and $Q_{y}$-PDW2 states.}
\end{figure}

\begin{figure}[H]
\begin{centering}
\includegraphics[width=0.7\textwidth]{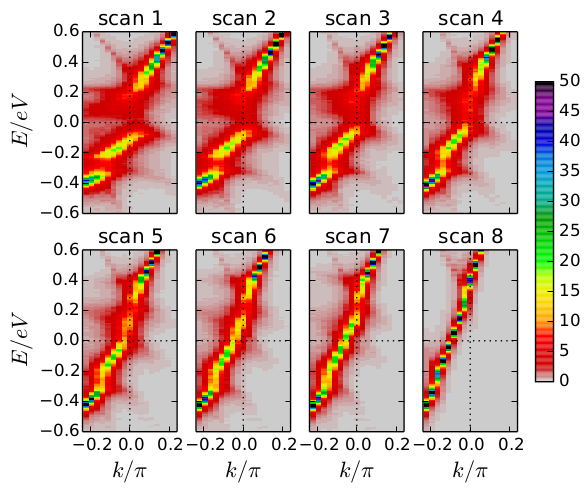}
\par\end{centering}

\caption{\label{fig:cmp-JS}Diagonal scans of spectral functions for the phase-separated
PDW4 state without thermal fluctuations {[}see Fig.\,10(a) in the
main text or Fig.\,\ref{fig:PDW4-SpecFun}(a) above{]}. The numbering
of the line scans corresponds to the red lines in Fig.\,\ref{fig:lineScans}.
Each line scan is in the direction away from the $\Gamma$ point and
the $k=0$ point coincides with the reduced Brillouin zone boundary.
The spectra should be compared with those in the ARPES by Yang \textit{et\,al.}\,\cite{Yang2008,PhysRevLett.107.047003}.
Firstly, the gap near the antinode is closed from below, a feature
that can be reproduced with PDW but not CDW order \cite{PhysRevX.4.031017}.
Secondly, the dispersion near the the end of the Fermi arc is not
particle-hole symmetric but bends downward and loses spectral weight.
Both features appear to be in qualitative agreement with the ARPES
results. Inclusion of random phases and thermal fluctuations seem
not affecting the dispersion except some broadening as we have seen
in Fig.\,\ref{fig:PDW4-SpecFun}, so we expect the results remain
valid as well for the PSPN state.}

\end{figure}

\begin{figure}[H]
\begin{centering}
\includegraphics[width=0.96\columnwidth]{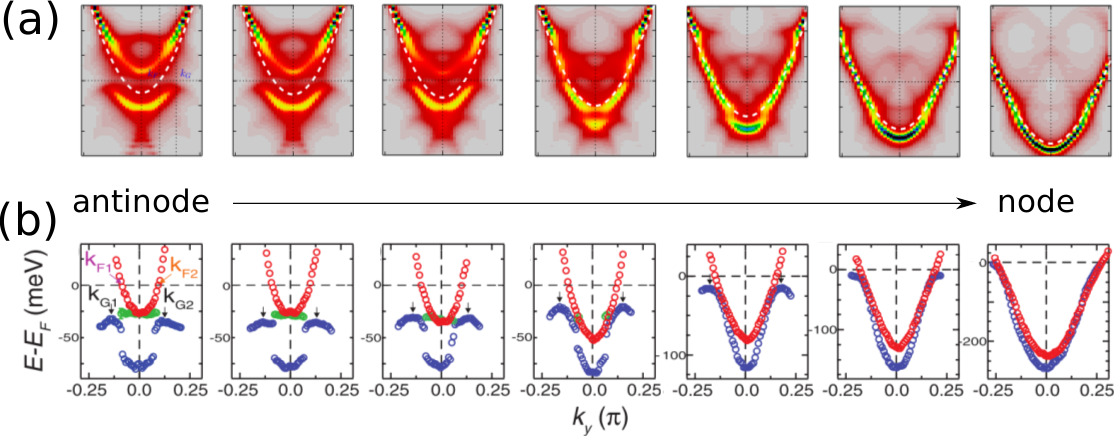}
\par\end{centering}

\caption{Comparison of the spectral functions for PSPN state and the ARPES
results. Panel (a) is taken from Fig.\,9(b) and shows the spectral
functions along different cuts in the BZ from antinode (left) to node
(right). The white dashed lines indicate the bare bands in different
BZ cuts. Panel (b) is taken from ARPES experiment \cite{He25032011}
(see also Ref.\,\cite{Hashimoto2010}) and shows the peaks of the
quasiparticle excitations. Blue and green circles are the peaks for
$T<T_{c}$ and red circles (bare band) are for $T>T^{*}$. The ARPES
results consist of denser cuts near the antinodal region, so we cannot
directly compare each pair of subgraphs. But we can compare the trend
and the details of the spectral functions. First, in both cases, the
spectrum is gapped near the antinode and becomes gapless near the
node, forming the so-called Fermi arc. Next, both show the presence
of the $k_{F}$-$k_{G}$ misalignment near the antinode. Also, the
gap near the antinode is closed by the spectral weight moving towards
the Fermi level from \textit{below} when moving away from the antinode
\cite{PhysRevX.4.031017,Wang2015a}, which is also captured by the
PSPN state in (a).}
\end{figure}